\documentclass[aps,superscriptaddress,preprintnumbers,reprint,amsmath,amssymb,floatfix,prl]{revtex4-2}

\usepackage[utf8]{inputenc}
\usepackage[T1]{fontenc}
\usepackage[colorlinks,allcolors=blue]{hyperref}
\usepackage{graphicx}
\usepackage{braket}

\begin{document}

\title{Anomalous diffusion in the Long-Range Haken-Strobl-Reineker model}

\author{A.~G.~Catalano}
\affiliation{University of Strasbourg and CNRS, CESQ and ISIS (UMR 7006), aQCess, 67000 Strasbourg, France}
\affiliation{Institut Ru\dj er Bo\v{s}kovi\'c, Bijeni\v{c}ka cesta 54, 10000 Zagreb, Croatia}

\author{F.~Mattiotti}
\affiliation{University of Strasbourg and CNRS, CESQ and ISIS (UMR 7006), aQCess, 67000 Strasbourg, France}

\author{J.~Dubail}
\affiliation{University of Strasbourg and CNRS, CESQ and ISIS (UMR 7006), aQCess, 67000 Strasbourg, France}
\affiliation{Universit\'{e} de Lorraine and CNRS, LPCT (UMR 7019), 54000 Nancy, France}

\author{D.~Hagenm\"uller}
\affiliation{University of Strasbourg and CNRS, CESQ and ISIS (UMR 7006), aQCess, 67000 Strasbourg, France}

\author{T.~Prosen}
\affiliation{Faculty for Mathematics and Physics, University of Ljubljana, Jadranska ulica 19, 1000 Ljubljana, Slovenia}

\author{F.~Franchini}
\affiliation{Institut Ru\dj er Bo\v{s}kovi\'c, Bijeni\v{c}ka cesta 54, 10000 Zagreb, Croatia}

\author{G.~Pupillo}
\affiliation{University of Strasbourg and CNRS, CESQ and ISIS (UMR 7006), aQCess, 67000 Strasbourg, France}
\affiliation{Institut Universitaire de France (IUF), 75000 Paris, France}

\begin{abstract}
    We analyze the propagation of excitons in a $d$-dimensional lattice with power-law hopping $\propto 1/r^\alpha$ in the presence of dephasing, described by a generalized Haken-Strobl-Reineker model. We show that in the strong dephasing (quantum Zeno) regime the dynamics is described by a classical master equation for an exclusion process with long jumps. In this limit, we analytically compute the spatial distribution, whose shape changes at a critical value of the decay exponent $\alpha_{\rm cr} = (d+2)/2$. The exciton always diffuses anomalously: a superdiffusive motion is associated to a L\'evy stable distribution with long-range algebraic tails for $\alpha\leq\alpha_{\rm cr}$, while for $\alpha > \alpha_{\rm cr}$  the distribution corresponds to a surprising mixed Gaussian profile with long-range algebraic tails, leading to the coexistence of short-range diffusion and long-range L\'evy-flights. In the many-exciton case, we demonstrate that, starting from a domain-wall exciton profile, algebraic tails appear in the distributions for any $\alpha$, which affects thermalization: the longer the hopping range, the faster equilibrium is reached. Our results are directly relevant to experiments with cold trapped ions,  Rydberg atoms and supramolecular dye aggregates. They provide a way to realize an exclusion process with long jumps experimentally.
\end{abstract}

\maketitle

\paragraph{Introduction.} Energy transport is of fundamental importance in biological, chemical, and physical systems. In light-harvesting setups, for example, solar energy is converted into excitons that are transported to a reaction center or to the interface between two different semiconductors, which often relies on long-range dipolar couplings between the excitons~\cite{doi:10.1146/annurev.physchem.54.011002.103746,photovoltaics_forster,C3EE42444H}. Transport then results from a competition between coherent hopping that tends to delocalize the wavefunctions and local couplings to vibrational, motional degrees of freedom and disorder potentials, which lead to the localization of carriers~\cite{doi:10.1098/rsif.2013.0901,PhysRev.109.1492,Kramer_1993,RevModPhys.80.1355}, limiting the conversion efficiency of optoelectronic devices~\cite{exciton_diffusion}. Theory has mostly focused on short-range couplings among quantum emitters, as they allow simple analytical approaches. For instance, the interplay between short-range hopping and local dephasing, which can be induced by, e.g., thermal noise or vibrational coupling~\cite{MUKAMEL1978327}, is captured by the Haken–Strobl–Reineker (HSR) model: for large enough dephasing, a transition from ballistic to diffusive motion occurs at time $t \sim 1/\gamma$~\cite{reineker,Haken,PhysRevLett.40.70}, with $\gamma$ the local dephasing rate. Diffusion taking place for $t \gg 1/\gamma$ is standard, i.e., an initially localized exciton spreads as a Gaussian distribution $\exp(-r^2/4Dt)$, with a diffusion coefficient $D=2J^2/\gamma$ ($J$ is the nearest neighbor hopping rate). While the HSR model with nearest-neighbor hopping has been extensively analyzed and even solved exactly~\cite{reineker,Haken,PhysRevLett.40.70,Moix_2013,medvedyeva2016exact}, the interplay of power-law long-range hopping and dephasing is more challenging and has not been analytically treated. Power-law hopping stems from the $\sim 1/r^3$ dipolar coupling in molecular aggregates~\cite{doi:10.1146/annurev.physchem.54.011002.103746,photovoltaics_forster,C3EE42444H} or nanocrystals~\cite{Mork2014,Chou2015,Penzo2020}, for instance, where large dephasing is naturally present~\cite{doi:10.1063/1.4944980,https://doi.org/10.1002/aenm.201700236,Becker2018,PRXQuantum.3.020354}. More general power-law-type couplings with arbitrary spatial decay can be engineered in artificial systems such as cold trapped ions~\cite{monroe2014,jurcevic2014} or Rydberg gases~\cite{PhysRevX.7.041063,doi:10.1126/science.aav9105}.

In this work, we investigate the HSR model with coupling between quantum emitters that decays with distance $r$ as a power-law $\sim 1/r^{\alpha}$, with variable power $\alpha$ and for a generic dimension $d$. In the presence of strong dephasing -- in the quantum Zeno regime ~\cite{PhysRevLett.122.050501} -- we map the system to a classical master equation (CME) that captures the long-time dynamics $t \gg 1/\gamma$, which we solve exactly by analytical and numerical means. 

\begin{figure}[t]
    \centering
    \includegraphics{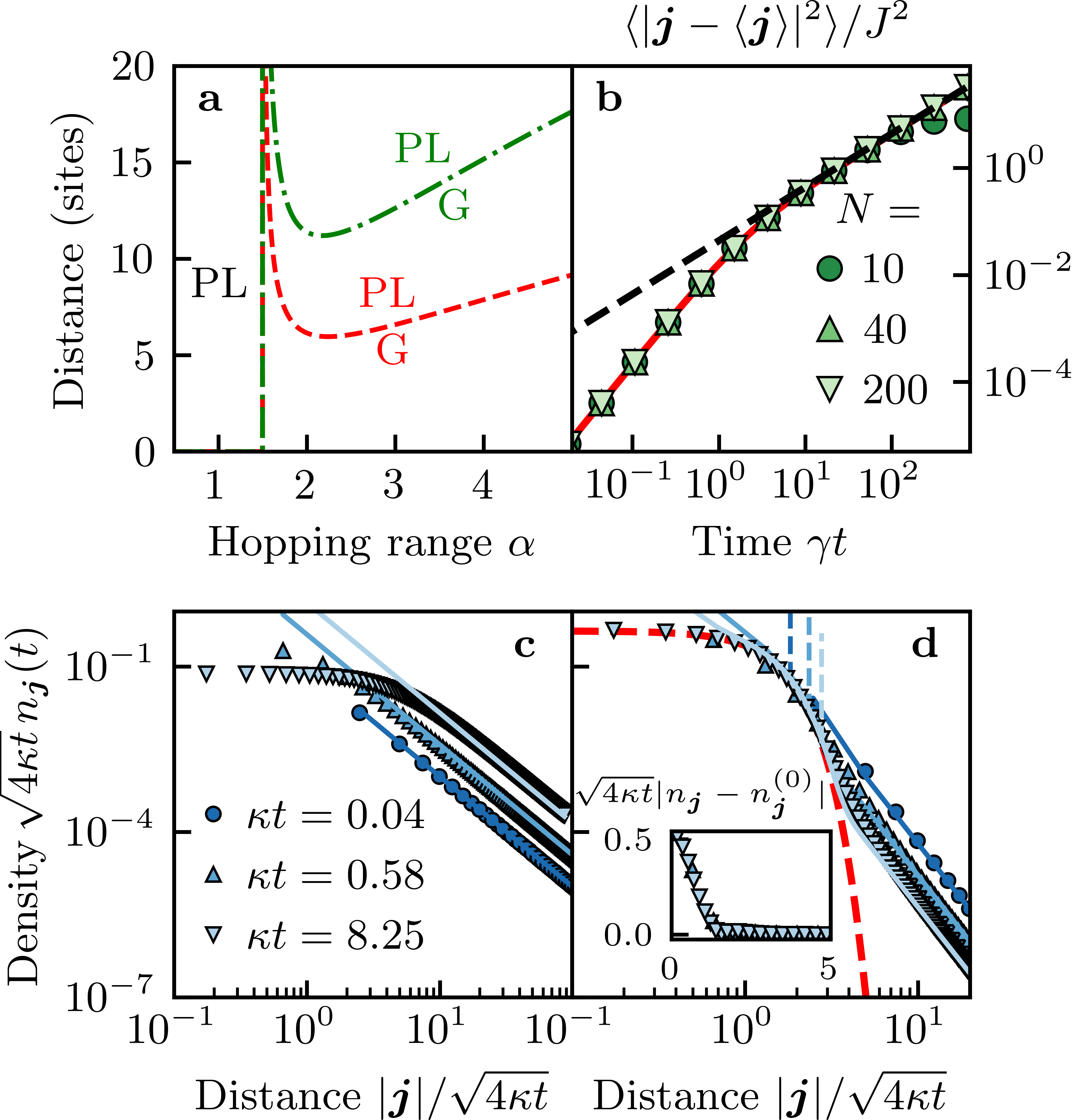}
    \caption{Single-exciton regime for $d=1$: time evolution of an exciton initially located at site $\boldsymbol{0}$. The exciton density profile $n_{\boldsymbol{j}}(t)$ is characterized by a power-law (PL) at long distance and a Gaussian (G) at short distance ({\bf a}). The boundary between the two regions (red dashed line for $\kappa t=1$ and green dashed-dotted line for $\kappa t = 3$) corresponds to $\xi_{\alpha,t}$ [see Eqs.~\eqref{eq:PLsummary}]. The quantum to classical crossover is illustrated through the time evolution of the exciton variance ({\bf b}), obtained by numerically solving Eq.~\eqref{eq:Lindblad} for $\alpha=3$ and $\gamma=10J$. Red solid line: exact solution Eq.~\eqref{eq:QtoC}, black dashed line: classical approximation for $\gamma t \gg 1$ [Eq.~\eqref{eq:QtoCapp}]. A pure power-law density profile for $\alpha=1 < \alpha_{\rm cr}$ ({\bf c}) and mixed Gaussian--power-law for $\alpha=2 > \alpha_{\rm cr}$ ({\bf d}) are obtained by numerically solving Eq.~\eqref{eq:creD} for $N=1000$ and $\gamma=10J$. Solid lines: approximation Eq.~\eqref{eq:PLsummary}, thick red dashed line: Gaussian term in Eq.~\eqref{eq:PLsummary2}, thin dashed lines: $\xi_{\alpha,t}$. The diffusion enhancement with respect to the case $\alpha \to \infty$ [$n_{\boldsymbol{j}}^{(0)}(t)$] is shown in the inset.}
    \label{fig:1D}
\end{figure}

We find that excitons always diffuse anomalously: in the single-exciton limit, the CME is the one of a discrete random walk with long jumps, or discrete L\'{e}vy flight~\cite{valdinoci2009long,dubkov2008levy,metzler2007some}, and for any finite $\alpha$ the exciton density profile always decays algebraically at long distances, in contrast to the standard diffusion obtained from the HSR model with nearest-neighbor hopping. The interaction range $\alpha$ determines whether the variance of the distribution converges or not: based on this, we define the critical exponent $\alpha_{\rm cr}=(d+2)/2$. For $\alpha \leq \alpha_{\rm cr}$, the dynamics is superdiffusive and the exciton density at sufficiently long distance is always a L\'evy stable distribution~\cite{zolotarev1999superdiffusion,janson2011stable,metzler2007some,dubkov2008levy} characterized by a long-range algebraic tail $\sim 1/r^{2\alpha}$. For $\alpha > \alpha_{\rm cr}$ and small enough time, the exciton density is also solely characterized by an algebraic tail, while at long time it exhibits a surprising mixed profile corresponding to a Gaussian distribution at short distance and an algebraic tail at large distance (Fig.~\ref{fig:1D}{\bf a}). The Gaussian part of the distribution mimics the standard diffusion in the HSR model. However, remarkably, also this Gaussian contribution is non-standard as the diffusion coefficient depends on $\alpha$ and is enhanced by the long range character of the hopping. We show that this finding is relevant to long-range exciton diffusion in light-harvesting systems such as nanocrystal quantum dots, where discrepancies between experimental observations and theory have been reported. 

We find that in the case of many excitons our model is equivalent to a long-jump symmetric exclusion process~\cite{jara2007non,jara2008hydrodynamic,bernardin2020hydrodynamic}, with a Markov matrix identical to the Hamiltonian of a long-range ferromagnetic Heisenberg model. Long-range hopping enhances exciton propagation so that equilibrium is reached faster as $\alpha$ is decreased. We capture the equilibration dynamics analytically via a continuous diffusion equation with fractional laplacian that qualitatively reproduces the numerical results for all $\alpha$. 


Excitons are modelled as spin-1/2 operators $S$. We start with the single-exciton case and study the dynamics in the presence of dephasing governed by the HSR quantum master equation
\begin{equation}
    \label{eq:Lindblad}
    \dot{\rho}=-i[H,\rho]+\gamma\sum_{\boldsymbol{j}}\bigg(L_{\boldsymbol{j}} \rho L_{\boldsymbol{j}}^\dagger - \frac{1}{2}\{L_{\boldsymbol{j}}^\dagger L_{\boldsymbol{j}},\rho\} \bigg)=\hat{\mathcal{L}}\rho.
\end{equation}
In our case, the coherent dynamics is described by the power-law hopping Hamiltonian
\begin{equation}
    H=\frac{1}{2}\sum_{\boldsymbol{j}}\sum_{\boldsymbol{r}\neq \boldsymbol{0}} \frac{J}{r^{\alpha}} \left(S^{+}_{\boldsymbol{j}} S^{-}_{\boldsymbol{j}+\boldsymbol{r}} + S^-_{\boldsymbol{j}} S^+_{\boldsymbol{j}+\boldsymbol{r}} \right),
    \label{Hamiltonian_LR}
\end{equation}
with $\rho$ the density matrix, $\boldsymbol{j}\in\mathbb{Z}^d$ the position in a $d$-dimensional lattice, $r=|\boldsymbol{r}|$, and $L_{\boldsymbol{j}}=L_{\boldsymbol{j}}^\dagger=S^{z}_{\boldsymbol{j}}$ the local dephasing operators, in the Lindblad formalism~\cite{breuerpetruccione,pichler2010nonequilibrium}. For $d=1$ and when a single exciton is initially present on a given site, it is known that the variance of the exciton evolves in time as~\cite{reineker}
\begin{align}
\label{eq:QtoC}
    \braket{|\boldsymbol{j}-\braket{\boldsymbol{j}}|^2} = 2\sum_{\boldsymbol{r}} \frac{r^2 H_{\boldsymbol{r}}^2}{\gamma^2} \left( \gamma t + e^{-\gamma t} - 1 \right),
\end{align}
with $H_{\boldsymbol{r}}=\bra{G}S^-_{\boldsymbol{j}} H S^+_{\boldsymbol{j}+{\boldsymbol{r}}}\ket{G}$ and $\ket{G}$ the ground state with all the spins down. The short- and long-time approximations of Eq.~\eqref{eq:QtoC} read
\begin{equation}
\label{eq:QtoCapp}
    \braket{|\boldsymbol{j}-\braket{\boldsymbol{j}}|^2} \approx
    \begin{cases}
        \sum_{\boldsymbol{r}} r^2 H_{\boldsymbol{r}}^2 t^2 & \text{for } \gamma t \ll 1 \\
        \displaystyle 2\sum_{\boldsymbol{r}} \frac{r^2 H_{\boldsymbol{r}}^2}{\gamma} t & \text{for } \gamma t \gg 1,
    \end{cases}
\end{equation}
and reveal a crossover in the dynamics: while a coherent quantum dynamics dominates for short time, a classical diffusive-like behaviour emerges for $t \gg 1/\gamma$. This is illustrated in Fig.~\ref{fig:1D}{\bf b}, where the exciton variance is obtained by numerically solving the quantum master equation~\eqref{eq:Lindblad} for different system sizes $N$, and compared to the analytical solutions Eqs.~\eqref{eq:QtoC} and \eqref{eq:QtoCapp}. The crossover from ballistic to diffusive regime is clearly visible. Interestingly, the transition to the classical regime always occurs at $t \sim 1/\gamma$, independently of $N$ and $\alpha$~\cite{Note3}. This is because in Eq.~\eqref{eq:QtoCapp}, the same multiplicative factor $\sum_{\boldsymbol{r}} r^2 H_{\boldsymbol{r}}^2$ governs both the short- and late-time behaviors, so the crossover time scale is independent of the details of the Hamiltonian. In Fig.~\ref{fig:1D}{\bf b} we see that for $\gamma t \gtrsim 10$, the quantum dissipative evolution is indistinguishable from the long-time asymptotics in Eq.~(\ref{eq:QtoCapp}).

Importantly, Eq.~\eqref{eq:QtoC} implies that the late-time diffusive-like regime is always reached, for any dephasing strength $\gamma$. This can also be seen from the QME~\eqref{eq:Lindblad}. Indeed, for any dephasing, we observe numerically that for large system size and long time ($t\gg 1/\gamma$) the coherences in the single-particle density matrix, $G_{\boldsymbol{j},\boldsymbol{m}}=\textrm{Tr}[\rho S^+_{\boldsymbol{j}}S^-_{\boldsymbol{m}}]$, with $\boldsymbol{j}\neq\boldsymbol{m}$, become negligible with respect to the population density $n_{\boldsymbol{j}}=G_{\boldsymbol{j},\boldsymbol{j}}$. However, in the limit of weak dephasing, this effect cannot simply be explained from perturbation theory in $\gamma$, as the long-time dynamics is determined by a non-perturbative branch of eigenmodes of the Liouvillian $\hat{\cal L}$ [Eq.~\eqref{eq:Lindblad}]~\cite{Note3}. An analogous effect has been observed in the case of nearest-neighbors hopping with dephasing and more sophisticated techniques should be used~\cite{medvedyeva2016exact}. We leave this for future work. Next we turn to the strong dephasing limit, which can be handled analytically more easily.

\paragraph{Strong dephasing: mapping to classical Markov process.} Following Refs.~\cite{Cai,Bernard_2018}, we 
use a second-order perturbative analysis, deriving an effective Liouvillian $\hat{\mathcal{L}}_{\rm eff}$ in the limit $\gamma \gg J$ (for similar treatments of the strong dissipative limit, see also~\cite{bernier2013emergence,poletti2013emergence} for soft-core bosons and nearest-neighbor hopping, or ~\cite{ciracNJP2009,rossini2021strong} for  atom losses instead of dephasing). We split the Liouvillian Eq.~(\ref{eq:Lindblad}) into two contributions, a term $\hat{\mathcal{L}}_0\rho=\gamma\sum_{\boldsymbol{j}}(S_{\boldsymbol{j}}^z\rho S_{\boldsymbol{j}}^z-\rho/4)$, and a perturbation $\hat{\mathcal{L}}_1\rho=-i[H,\rho]$. We find~\cite{Note3} that the effective dynamics $\dot{\rho} = \mathcal{L}_{\rm eff}  \rho$ is governed by a CME for the probability distribution 
\begin{equation}
\label{eq:mbcl}
    \dot{p} (\boldsymbol{\sigma}) =  -\sum_{\boldsymbol{\sigma}'}  \left< \boldsymbol{\sigma} \right| R \left| \boldsymbol{\sigma}' \right> p(\boldsymbol{\sigma}'),
\end{equation}
with $\ket{\boldsymbol{\sigma}}$ the eigenstates of the $S_{\boldsymbol{j}}^z$ operators, and $p(\boldsymbol{\sigma})$ the probability distribution defined by the diagonal entries of the density matrix $\rho = \sum_{\boldsymbol{\sigma}} p(\boldsymbol{\sigma}) \ket{\boldsymbol{\sigma}}\bra{\boldsymbol{\sigma}}$. The generator of the CME (\ref{eq:mbcl}) is that of an exclusion process with long jumps, which turns out to be identical to the following Hamiltonian of a long-range ferromagnetic Heisenberg model
\begin{align}
    \label{eq:K}
    R=-\sum_{\boldsymbol{j};\boldsymbol{r}\neq \boldsymbol{0}}\frac{2J^2}{\gamma r^{2\alpha}}\bigg[&\frac{1}{2}(S^+_{\boldsymbol{j}} S^-_{\boldsymbol{j}+\boldsymbol{r}}+S^-_{\boldsymbol{j}} S^+_{\boldsymbol{j}+\boldsymbol{r}}) \nonumber \\
    &+S_{\boldsymbol{j}}^zS_{\boldsymbol{j}+\boldsymbol{r}}^z-\frac{1}{4}\bigg].
\end{align}
A similar observation was made in Refs.~\cite{Cai,Bernard_2018} for strictly short-range models, whose strong-dephasing limit corresponds to a ferromagnetic Heisenberg model with short-range couplings; here we extend this result to long-range hopping. We note that, interestingly, the case $\alpha = d= 1$ in Eq.~\eqref{eq:K} corresponds to the Haldane-Shastry Hamiltonian~\cite{PhysRevLett.60.635,*PhysRevLett.60.639}, a famous quantum integrable model. For any exciton number, the associated exclusion process should then be exactly solvable by Bethe Ansatz techniques, which we will investigate in a future work.

\paragraph{Anomalous diffusion of single exciton.} We first focus on the classical dynamics dictated by Eq.~\eqref{eq:mbcl} for the case of a single exciton. Equation~\eqref{eq:K} provides the evolution of the population density 
\begin{equation}
    \label{eq:creD}
    \dot{n}_{\boldsymbol{j}}=\sum_{\boldsymbol{r}\neq \boldsymbol{0}}\frac{\kappa}{r^{2\alpha}} (n_{\boldsymbol{j}+\boldsymbol{r}}-n_{\boldsymbol{j}}),
\end{equation}
with the effective Zeno-like rate $\kappa=2J^2/\gamma$. An alternative derivation of Eq.~\eqref{eq:creD} is obtained by adiabatically eliminating the coherences of the single-exciton density matrix 
$G_{\boldsymbol{j},\boldsymbol{m}}$%
~\cite{reineker,PhysRevLett.122.050501,Note3}. Notice that Eq.~(\ref{eq:creD}) is well defined in the thermodynamic limit only if $\alpha > d/2$ so that $\sum_{\boldsymbol{r} \neq \boldsymbol{0}} r^{-2 \alpha}$ is finite. In order to solve Eq.~\eqref{eq:creD} for an exciton initially at the origin, $n_{\boldsymbol{j}}(t=0) = \delta_{\boldsymbol{j},\boldsymbol{0}}$, we introduce the characteristic function $K(\boldsymbol{q},t)=\sum_{\boldsymbol{j}}n_{\boldsymbol{j}}(t)e^{i\boldsymbol{q}\cdot \boldsymbol{j}}$, where $\boldsymbol{q}\in \mathbb{R}^d$. Using Eq.~\eqref{eq:creD}, we find that the characteristic function at time $t$ then reads 
\begin{equation}
    \label{eq:genfunc}
    K(\boldsymbol{q},t)=e^{(\mathcal{A}_{2\alpha,d}(\boldsymbol{q})-\mathcal{A}_{2\alpha,d}(\boldsymbol{0})) t},
\end{equation}
with the initial condition $K(\boldsymbol{q},0) = 1$, and $\mathcal{A}_{2\alpha,d}(\boldsymbol{q})=\kappa\sum_{\boldsymbol{r}\neq\boldsymbol{0}}r^{-2\alpha}e^{-i\boldsymbol{q}\cdot \boldsymbol{r}}$. Equation~\eqref{eq:genfunc} provides the time evolution of the mean position $\langle \boldsymbol{j}\rangle=-i\nabla_{\boldsymbol{q}}K(\boldsymbol{0},t)=\boldsymbol{0}$ and of the variance $\langle |\boldsymbol{j}|^2 \rangle=-\Delta_{\boldsymbol{q}}K(\boldsymbol{0},t)=2D_\alpha t$. The diffusion coefficient $D_\alpha=\frac{1}{2}\mathcal{A}_{2\alpha-2,d}(\boldsymbol{0})$ provides a first insight into the character of the dynamics for different $\alpha$ (however, see also discussion below): diffusive-like spreading of excitons takes place when $D_\alpha$ converges in the thermodynamic limit, which is ensured when $\alpha>\alpha_{\rm cr}$~\cite{Note3}. This corresponds to the quantum master equation solution in the regime $\gamma t \gg 1$, shown in Eq.~\eqref{eq:QtoCapp} and Fig.~\ref{fig:1D}{\bf b}. On the other hand, for $\alpha \leq \alpha_{\rm cr}$, $D_\alpha$ diverges and the dynamics is superdiffusive. Equation~\eqref{eq:genfunc} further allows one to determine the exciton density profile $n_{\boldsymbol{j}}(t)$ for all $\alpha$ and times $t$. Since the long-distance behavior of $n_{\boldsymbol{j}}(t)$ is determined by the singularity of $K(\boldsymbol{q},t)$ when $q \equiv |\boldsymbol{q}| \rightarrow 0$, we analyze $\mathcal{A}_{2\alpha,d}(\boldsymbol{q})$ in that limit. We find $\mathcal{A}_{2\alpha,d}(\boldsymbol{q})\approx\mathcal{A}_{2\alpha,d}(\boldsymbol{0})- C_\alpha q^{2\alpha-d}$ if $\alpha\le\alpha_{\rm cr}$, and $\mathcal{A}_{2\alpha,d}(\boldsymbol{q})\approx\mathcal{A}_{2\alpha,d}(\boldsymbol{0})-\frac{\mathcal{A}_{2\alpha-2,d}(\boldsymbol{0})}{2}q^2 - C_\alpha q^{2\alpha-d}$ if $\alpha>\alpha_{\rm cr}$~\cite{Note3}, with $C_\alpha = -\kappa\pi^{\frac{d}{2}}2^{d-2\alpha} \Gamma\big(\frac{d}{2}-\alpha\big)/ \Gamma\big(\alpha\big)$. The expression of $C_\alpha$ depends on the boundary conditions: here we have assumed translational invariance. Inserting these expressions into Eq.~(\ref{eq:genfunc}), the characteristic function finally reads
\begin{equation}
    K(\boldsymbol{q},t) \underset{q \rightarrow 0}{\simeq}  \left\{ \begin{array}{lll}
        e^{-C_\alpha q^{2\alpha -d} t  } && \alpha \leq \alpha_{\rm cr} \\   
        e^{-D_\alpha q^2 t} \, e^{-C_\alpha q^{2\alpha -d} t } && \alpha > \alpha_{\rm cr} .
    \end{array}
    \right. 
    \label{characteristic_fun}
\end{equation}
For $\alpha \leq \alpha_{\rm cr}$, this is the characteristic function of a L\'evy stable distribution~\cite{zolotarev1999superdiffusion,janson2011stable,metzler2007some,dubkov2008levy}, which is characterized by a long-range algebraic tail. Such a distribution corresponds to large but infrequent steps, the so-called rare events or big jumps relevant to a large variety of phenomena including motion of cold atoms in laser cooling, transport in turbulent flow, and neural transmission~\cite{Vezzani_rare_events}. For $\alpha > \alpha_{\rm cr}$, instead, the characteristic function has a peculiar mixed nature: it is the product of a Gaussian and of the L\'evy flight factor.

From the inverse Fourier transform of $K(\boldsymbol{q},t)$ we obtain the population $n_{\boldsymbol{j}}(t)$. 
For $\alpha \leq \alpha_{\rm cr}$ the asymptotic behavior $n_{\boldsymbol{j}}(t)$ depends on $\boldsymbol{j}$ as
\begin{subequations}
\label{eq:PLsummary}
\begin{equation}
\label{eq:PLsummary1}
    n_{\boldsymbol{\boldsymbol{j}}}(t) \underset{ |\boldsymbol{j} | \gg 1   }{\simeq}
    \kappa t/|\boldsymbol{j}|^{2\alpha} ,
\end{equation}
while for $\alpha > \alpha_{\rm cr}$ we obtain the following mixed Gaussian and power-law behavior with increasing $|\boldsymbol{j}|$ 
\begin{equation}
\label{eq:PLsummary2}
    n_{\boldsymbol{\boldsymbol{j}}}(t) \simeq \begin{cases}
        \displaystyle  \frac{\exp(-|\boldsymbol{j}|^2/4D_\alpha t)}{(4\pi D_\alpha t)^{d/2}} & | \boldsymbol{j} | \lesssim \xi_{\alpha,t}  \\
    \displaystyle 
    \kappa t/|\boldsymbol{j}|^{2\alpha}  & |
    \boldsymbol{j} |  \gg  \xi_{\alpha,t}  ,
    \end{cases}
\end{equation}
\end{subequations}
which is one of the main results of this work. In Eq.~(\ref{eq:PLsummary2}), $\xi_{\alpha,t}$ is the length scale at which the behavior crosses over from Gaussian to power-law. For large enough time, $\xi_{\alpha,t}$ is well approximated by $\xi_{\alpha,t} \approx \sqrt{4D_\alpha t \log [4\alpha^\alpha \pi^{-d/2} \kappa^{-1} D_\alpha(4 D_\alpha t)^{\alpha-\alpha_{\rm cr}}]}$~\cite{Note3}. The exact expression of $\xi_{\alpha,t}$ exhibits a minimum as a function of $\alpha$, and a discontinuity at $\alpha=\alpha_{\rm cr}$ [Fig.~\ref{fig:1D}{\bf a}]. For large $\alpha$, $\xi_{\alpha,t}$ increases with $\alpha$ as $\xi_{\alpha,t}\sim \sqrt{4D_\alpha t \alpha \log \alpha}$, and we ultimately recover a standard diffusive (Gaussian) behavior for $\alpha \to \infty$. For $\alpha \to \alpha_{\rm cr}^+$, $D_\alpha$ diverges and therefore $\xi_{\alpha,t}$ does too. For small enough time, the power-law behavior takes over for all $\alpha$. We emphasize that since $\xi_{\alpha,t}$ grows with time, the Gaussian dynamics ultimately dominates at long times for $\alpha > \alpha_{\rm cr}$, and thus we expect the algebraic tail to particularly affect transient phenomena. 

This behavior is illustrated in Fig.~\ref{fig:1D}{\bf c},{\bf d} for $d=1$, where we show a numerical solution of the CME~\eqref{eq:creD} together with the asymptotic behavior Eq.~\eqref{eq:PLsummary}. For $\alpha< \alpha_{\rm cr}$, the distribution is only characterized by a power-law decay with amplitude growing linearly with time and independent of the lattice dimension $d$ [Fig.~\ref{fig:1D}{\bf c}]. The scaling with the distance $1/|\boldsymbol{j}|^{2\alpha}$ turns out to be the same as the hopping rate. While the decay of the distribution still goes as $\sim1/|\boldsymbol{j}|^{2\alpha}$ at long distances for $\alpha > \alpha_{\rm cr}$, diffusion dominates at short distances showing a Gaussian profile [Fig.~\ref{fig:1D}{\bf d}], but with an enhanced diffusion coefficient $D_\alpha$ as compared to the nearest-neighbor case (inset). In the usual dipolar coupling case $\alpha=d=3$, for instance, we find that $D_{\alpha}$ is enhanced by a factor $\approx 2.8$ as compared to standard diffusion with nearest-neighbor hopping. Interestingly, we find that those power-law tails have a profound effect on the dynamics in the presence of strong dephasing for all $\alpha$, which is surprising for $\alpha > \alpha_{\rm cr}$ where a simple diffusive behavior is expected from short-range models~\cite{medvedyeva2016exact}. In the following, we illustrate this effect for the case of many excitons following a quench.

\begin{figure}[t]
    \centering
    \includegraphics{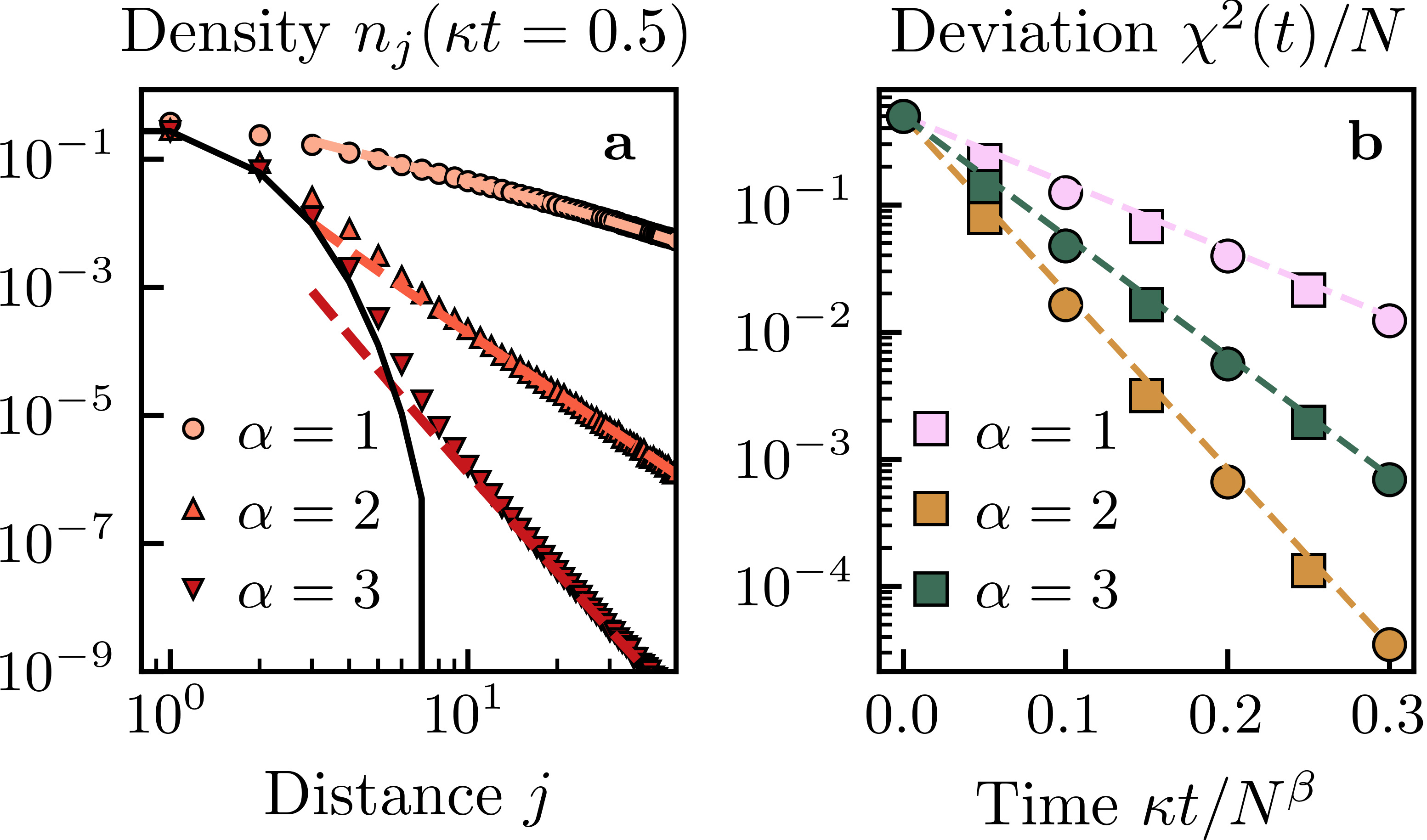}
    \caption{Speedup of the relaxation dynamics for $d=1$. Starting from a domain-wall exciton profile, the occupation profile $n_j(t)$ is computed numerically from Eq.~\eqref{eq:mbcl} for $N=100$ and $\kappa t = 0.5$ ({\bf a}), and exhibits power-law tails showing that equilibrium is reached faster as $\alpha$ is decreased. The continuous line corresponds to nearest-neighbour hopping, and the dashed lines to the approximate solution Eq.~\eqref{occup_pro}. {\bf b} Time evolution of the deviation from equilibrium $\chi^2(t)$ for different $\alpha$ and $N$. The circles and squares are for $N=100$ and $N=1000$, respectively. The dashed lines are the best fit $\propto\exp(-t/\tau)$, with $\tau$ given by Eq.~\eqref{eq:tau}.}
    \label{fig:mb}
\end{figure}
\paragraph{Many excitons: speedup of relaxation.} We consider the dynamics in the many-exciton sector of Eq.~\eqref{eq:mbcl} on a $d=1$ lattice, starting from a ``domain-wall'' initial condition, where the leftmost $N/2$ sites are all occupied, while the other sites are empty, in analogy with a Joule expansion. We analyze the occupation profile at time $t$, i.e. $n_j(t)={\rm Tr}[\rho(t)S^+_jS^-_j]$, where $\rho(t)$ is the density matrix solving Eq.~\eqref{eq:mbcl}. Both for $\alpha<\alpha_{\rm cr}$ and for $\alpha>\alpha_{\rm cr}$, a flat equilibrium solution is reached at large $t$~\cite{Note3}, such that $\Bar{n}=\lim_{t\to\infty}n_j(t)=0.5$ $\forall j$. Interestingly here, the equilibrium is reached for any hopping range $\alpha$, which is in contrast to the purely quantum case, where long-range interactions can break ergodicity in the absence of disorder~\cite{santos2022,santos2021,rigol2013}.

For short time $\kappa t \ll N^{2\alpha}$, the distribution away from the origin is dominated by single exciton hopping events, and we find that the profile has power-law tails~\cite{Note3} 
\begin{equation}
n_j(t)\propto \kappa t \int_{-N/2}^0 (j+r)^{-2\alpha}dr \approx \kappa t / j^{2\alpha-1},
\label{occup_pro}
\end{equation}
as shown in Fig.~\ref{fig:mb}{\bf a}. As a consequence, the exciton spreads faster as $\alpha$ is decreased. To quantify how fast the equilibrium profile is reached, we compute the normalized chi-squared $\chi^2(t)/N=\sum_j[n_j(t)-\bar{n}]^2/(N\bar{n})$ between the profile at time $t$ and the equilibrium one. Figure~\ref{fig:mb}{\bf b} shows that the equilibrium regime is reached exponentially in time for any $\alpha$, $\chi^2(t)/N \propto \exp(-t/\tau)$. Note that this scaling can be recovered by analyzing the gap of the Liouvillian, Eq.~\eqref{eq:mbcl}, which follows from the spinon dispersion of the ferromagnetic Heisenberg model~\cite{nakano1994quantum}. We observe that the half-time of the exponential increases with a power of the system size $N$ as
\begin{equation}
    \label{eq:tau}
    \tau = \frac{N^\beta}{2 \pi^\beta b_\alpha} \qquad \text{with} \qquad
    \beta=
    \begin{cases}
    2\alpha-1 & \alpha < \alpha_{\rm cr} \\
    2 & \alpha > \alpha_{\rm cr}
    \end{cases}
   ~,
\end{equation}
for some constant $b_\alpha$, while $\tau = \frac{N^2 \log N}{2\pi^2 b_\alpha}$ in the critical case $\alpha=\alpha_{\rm cr}=3/2$. Notice that the scaling (\ref{eq:tau}) is precisely what is expected from the continuous diffusion equation with (fractional) Laplacian,
\begin{equation}
    \label{eq:fracdiff}
    \frac{\partial n(x,t)}{\partial t} = b_\alpha \Delta^{\beta/2}n(x,t).
\end{equation}
Indeed, the solution to this evolution equation with an initial domain-wall density profile has the Fourier decomposition $n(x,t)=\frac{1}{2}+\sum_{m\in\mathbb{N}} c_m(t) \cos(\pi m x/N)$ with coefficients decaying as $c_m(t)\propto\exp(-b_\alpha (m\pi/N)^\beta t)$, thus $\chi^2(t) \propto N [n(x,t)-1/2]^2 \propto \exp(-2 \pi^{\beta} b_{\alpha} t /N^{\beta})$.

The fact that the large-scale evolution of our system should be captured by a continuous diffusion equation with fractional Laplacian (\ref{eq:fracdiff}) follows from the form of the generator of the CME (\ref{eq:K}), which is SU(2) symmetric. Indeed, exploiting the SU(2) symmetry, one can switch from one `magnetization sector' to another ---{\it i.e.} from one exciton number to another--- without changing its spectrum. This suggests that the equation governing the evolution of the density profile for many excitons at large scales should be the same as for a single exciton. In particular, the constant $b_\alpha$ in Eq.~(\ref{eq:tau}) is expected to match the diffusion constant of a single exciton, i.e. $b_\alpha=D_\alpha$ for $\alpha>\alpha_{\rm cr}$ and $b_{\alpha}=C_\alpha$ for $\alpha<\alpha_{\rm cr}$. From the data in Fig.~\ref{fig:mb}, we find the numerical values $b_{\alpha}/\kappa \simeq 1.93,1.62,1.1$ for $\alpha=1,2,3$, to be compared with the analytical result $C_1/\kappa=3.14$, $D_2/\kappa = 1.64$, $D_3/\kappa =1.08$. The agreement is very good for $\alpha > \alpha_{\rm cr}$, however the values differ in the long-range case $\alpha < \alpha_{\rm cr}$: this discrepancy is due to the different boundary conditions between the numerics in Fig.~\ref{fig:mb} (open boundary conditions) and in the analytical derivation of $C_\alpha$ (which assumes translational invariance, {\it i.e.} periodic boundary conditions). We also emphasize that $\beta$ decreases with $\alpha$ for $\alpha<\alpha_{\rm cr}$, which implies that the equilibrium is reached faster (for large $N$) as the interaction range increases.

\paragraph{Outlook.}
Our results provide a way to experimentally realize an exclusion process with long jumps~\cite{jara2007non,jara2008hydrodynamic,bernardin2020hydrodynamic}, and are highly relevant to nanocrystal quantum dots that are attracting more and more interest for solar cell applications~\cite{Becker2018}. In particular, discrepancies between the exciton diffusion length measured experimentally and the values predicted by standard diffusion theory applied to F\"orster energy transfer ($\alpha=3$) have been recently reported~\cite{Giovanni2021,Mork2014}. We argue in the supplemental material that such discrepancies would typically be reduced by a factor of $\sim 2$ upon properly including the long range character of the hopping in the diffusion coefficient, which is not the case in standard diffusion models assuming nearest-neighbor hopping~\cite{exciton_diffusion}. Our model is also relevant to molecular aggregates that play an important role in photosynthetic complexes and optoelectronic devices~\cite{Wurthner2011}. Dye monomers interacting via dipole-dipole coupling ($\alpha=3$) can indeed form highly-ordered assemblies~\cite{doi:10.1021/acs.chemrev.7b00581}. Supramolecular chemistry offers the possibility to control the mutual arrangement of monomers to achieve a nearest-neighbor hopping $J <3~{\rm THz}$, while the typical dephasing rate can exceed $14~{\rm THz}$ at room temperature~\cite{https://doi.org/10.1002/aenm.201700236,doi:10.1063/1.4944980}. Our model could also be realized with ions in linear Paul traps, with $J\approx 100-1000$~Hz and the possibility to tune the hopping range within $0<\alpha<3$~\cite{jurcevic2014,RevModPhys.93.025001,PRXQuantum.2.020343}. Controlled dephasing can be realized via detuned lasers that induce time-dependent ac-Stark shifts~\cite{PhysRevA.97.023606}, allowing to reach the large dephasing regime with $\gamma>10J$~\cite{PhysRevLett.122.050501}. A similar implementation could also be achieved with Rydberg atoms~\cite{PhysRevLett.115.093002}, where the $\gamma\gg J$ regime can be reached for large atom densities.

\footnotetext[3]{See supplemental material including plots of the exciton density profile for $d >1$, an alternative derivation of the CME Eq.~\eqref{eq:creD}, the full derivation of: $\braket{\boldsymbol{j}(t)}$ and $\braket{|\boldsymbol{j}|^2(t)}$, $\alpha_{\rm cr}$, ${\cal A}_{2\alpha,d}(\boldsymbol{q})$, $n_{\boldsymbol{j}}(t)$, $\xi_{\alpha,t}$, Eq.~\eqref{eq:K}, the many-exciton $n_j(t)$, as well as a treatment of the weak dephasing regime. The supplemental material includes Refs.~\cite{nist,Cai,reineker,Giovanni2021,Mork2014,eisler2011crossover}.}

\paragraph{Acknowledgements.} We thank Shannon Whitlock, Johannes Schachenmayer and Philipp Hauke for stimulating discussions. Numerical code for this work has been written in \emph{Julia}~\cite{shahSR2017} and \emph{Python}, using \emph{QuantumOptics.jl}~\cite{ritschCPC2018} for the quantum master equation simulations. A.G.C. acknowledges support from the MOQS ITN programme, a European Union’s Horizon 2020 research and innovation program under the Marie Skłodowska-Curie grant agreement number 955479. F.F. acknowledges support from the Croatian Science Funds Project No. IP-2019–4–3321 and the QuantiXLie Center of Excellence, a project co-financed by the Croatian Government and European 12 Union through the European Regional Development Fund–the Competitiveness and Cohesion Operational Programme (Grant KK.01.1.1.01.0004). F.M. and G.P. acknowledge funding from the French ANR via project CLIMAQS. Computing time was provided by the High-Performance Computing Center of the University of Strasbourg.

\bibliographystyle{apsrev4-2}
\bibliography{ref}

\end{document}


\renewcommand{\theequation}{S\arabic{equation}} 
\renewcommand{\thefigure}{S\arabic{figure}} 

\title{Supplemental Material for ``Anomalous diffusion in the Long-Range Haken-Strobl-Reineker model''}

\author{A.~Catalano}
\affiliation{University of Strasbourg and CNRS, CESQ and ISIS (UMR 7006), aQCess, 67000 Strasbourg, France}%
\affiliation{Institut Ru\dj er Bo\v{s}kovi\'c, Bijeni\v{c}ka cesta 54, 10000 Zagreb, Croatia}

\author{F.~Mattiotti}
\affiliation{University of Strasbourg and CNRS, CESQ and ISIS (UMR 7006), aQCess, 67000 Strasbourg, France}%

\author{J.~Dubail}
\affiliation{University of Strasbourg and CNRS, CESQ and ISIS (UMR 7006), aQCess, 67000 Strasbourg, France}%
\affiliation{Universit\'{e} de Lorraine and CNRS, LPCT (UMR 7019), 54000 Nancy, France}

\author{D.~Hagenm\"uller}
\affiliation{University of Strasbourg and CNRS, CESQ and ISIS (UMR 7006), aQCess, 67000 Strasbourg, France}%

\author{T.~Prosen}
\affiliation{Faculty for Mathematics and Physics, University of Ljubljana, Jadranska ulica 19, 1000 Ljubljana, Slovenia}

\author{F.~Franchini}
\affiliation{Institut Ru\dj er Bo\v{s}kovi\'c, Bijeni\v{c}ka cesta 54, 10000 Zagreb, Croatia}

\author{G.~Pupillo}
\affiliation{University of Strasbourg and CNRS, CESQ and ISIS (UMR 7006), aQCess, 67000 Strasbourg, France}%
\affiliation{Institut Universitaire de France (IUF), 75000 Paris, France}

\maketitle

In this supplemental material we provide an alternative derivation of the CME Eq.~(7), and the full derivation of: the moments $\braket{\boldsymbol{j}(t)}$ and $\braket{|\boldsymbol{j}|^2(t)}$, the critical exponent $\alpha_{\rm cr}$, the approximate function ${\cal A}_{2\alpha,d}(\boldsymbol{q})$, the exciton density profile $n_{\boldsymbol{j}}(t)$, the length scale $\xi_{\alpha,t}$, the effective Liouvillian Eq.~(6), and the approximate occupation profile $n_j(t)$ in the many-body case. We also provide a treatment of the weak dephasing regime.

\section{The single-particle problem}

\subsection{From the quantum master equation (QME) to the classical master equation (CME): alternative derivation}\label{sec:QME-CRE}

We consider a long-range spin-$\frac{1}{2}$ model on a $d$-dimensional hypercubic lattice $\mathbb{Z}^d$ in presence of pure dephasing through the Lindblad QME
\begin{equation}
    \dot{\rho}=-i[H,\rho]+\gamma\sum_{\boldsymbol{j}}\bigg(L_{\boldsymbol{j}}^\dagger \rho L_{\boldsymbol{j}} - \frac{1}{2}\{L_{\boldsymbol{j}}L_{\boldsymbol{j}}^\dagger,\rho\} \bigg),
\end{equation}
with $L_{\boldsymbol{j}}=S^z_{\boldsymbol{j}}$, and the Hamiltonian given by
\begin{equation}
    \label{eq:Hamiltonian}
    H=\frac{J}{2}\sum_{\boldsymbol{j}} \sum_{\boldsymbol{r}\neq\boldsymbol{0}}|\boldsymbol{r}|^{-\alpha}(S^+_{\boldsymbol{j}} S^-_{\boldsymbol{j}+\boldsymbol{r}} + h.c.).
\end{equation}
In particular, we study the evolution of the two-point correlation functions $G_{\boldsymbol{j},\boldsymbol{m}}=\Tr[\rho S^+_{\boldsymbol{j}} S^-_{\boldsymbol{m}}]$. We focus on the single particle subspace, spanned by the states $\{\ket{\boldsymbol{j}}\}$ representing a configuration in which all spins are down but at the lattice site $\boldsymbol{j}$. 

Within this subspace the two-point correlation functions evolve according to
\begin{equation}
    \label{eq:Coherences}
    \dot{G}_{\boldsymbol{j},\boldsymbol{m}}=iJ\sum_{\boldsymbol{r}\neq \boldsymbol{0}}(G_{\boldsymbol{j}+\boldsymbol{r},\boldsymbol{m}}-G_{\boldsymbol{j},\boldsymbol{m}+\boldsymbol{r}})|\boldsymbol{r}|^{-\alpha} - \gamma G_{\boldsymbol{j},\boldsymbol{m}},
\end{equation}
\begin{equation}
    \label{eq:population}
    \dot{G}_{\boldsymbol{j},\boldsymbol{j}}=iJ\sum_{\boldsymbol{r}\neq\boldsymbol{0}}(G_{\boldsymbol{j}+\boldsymbol{r},\boldsymbol{j}}-G_{\boldsymbol{j},\boldsymbol{j}+\boldsymbol{r}})|\boldsymbol{r}|^{-\alpha}.
\end{equation}
Following the idea of \cite{reineker}, assuming $\dot{G}_{\boldsymbol{j},\boldsymbol{m}}\ll \gamma G_{\boldsymbol{j},\boldsymbol{m}}$, i.e. that the phase relations are destroyed very rapidly, we can neglect the time derivative in \eqref{eq:Coherences}. Moreover, considering only time intervals larger than the decay time of the phase, we may also neglect non-diagonal terms w.r.t. to diagonal ones. Then, we obtain an expression for $G_{\boldsymbol{j},\boldsymbol{m}}$ in terms of the diagonal elements:
\begin{equation}
    \label{eq:Coh2}
    G_{\boldsymbol{j},\boldsymbol{m}}=\frac{iJ}{\gamma}(G_{\boldsymbol{m},\boldsymbol{m}}-G_{\boldsymbol{j},\boldsymbol{j}})|\boldsymbol{j}-\boldsymbol{m}|^{-\alpha}.
\end{equation}
We can substitute the latter expression into \eqref{eq:population}, and arrive to the following equation for the diagonal elements
\begin{equation}
    \dot{G}_{\boldsymbol{j},\boldsymbol{j}}=\frac{2J^2}{\gamma}\sum_{\boldsymbol{r}\neq\boldsymbol{0}}(G_{\boldsymbol{j}+\boldsymbol{r},\boldsymbol{j}+\boldsymbol{r}}-G_{\boldsymbol{j},\boldsymbol{j}})|\boldsymbol{r}|^{-2\alpha}.
\end{equation}
The latter expression can be recognized as the classical master equation (CME)~(7) for a random walker in a $d$-dimensional hypercubic lattice with long-range hopping.  
In particular, with hopping rate from site $\boldsymbol{j}$ to site $\boldsymbol{m}$ given by $\kappa|\boldsymbol{j}-\boldsymbol{m}|^{-2\alpha}$, then the CME for the probability $n_{\boldsymbol{j}}(t)$ of finding the walker at site $\boldsymbol{j}$ at time $t$ reads
\begin{equation}
    \label{eq:RE1}
    \dot{n}_{\boldsymbol{j}}=\kappa \sum_{\boldsymbol{r}\neq 0}(n_{\boldsymbol{j}+\boldsymbol{r}}-n_{\boldsymbol{j}})|\boldsymbol{r}|^{-2\alpha},
\end{equation}
where the classical rate $\kappa$ is related to the parameters entering the quantum master equation by 
\begin{equation}
\kappa=\frac{2J^2}{\gamma}.
\end{equation}


\begin{figure}[!tb]
    \centering
    \includegraphics{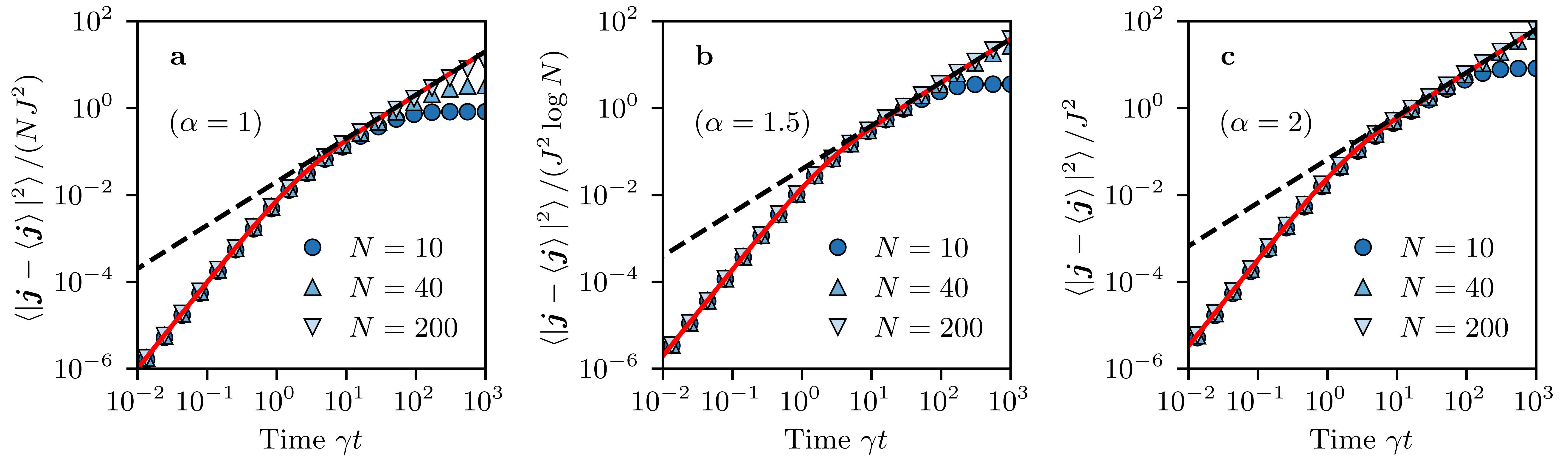}
    \caption{Variance of the excitation as a function of time, for a $d=1$ array of $N$ emitters, determined by numerically solving the QME~(1). In all panels $\gamma=10J$, the red continuous lines represent Eq.~(3), while the black dashed lines are the classical approximations, see Eq.~(4). As discussed in the text, the variance is normalized in {\bf a} ($\alpha=1<\alpha_{\rm cr}$) and {\bf b} ($\alpha=\alpha_{\rm cr}=3/2$) in order to avoid its divergence in the $N\to\infty$ limit.
    }
    \label{fig:DeltaX2}
\end{figure}

The quantum-to-classical transition is illustrated in Fig.~1{\bf b} of the main text, where the variance of an initially localized excitation is plotted as a function of time, for $d=1$ and $\alpha>\alpha_{\rm cr}$.
Here, in Fig.~\ref{fig:DeltaX2}{\bf a},{\bf b}, we report the results for $\alpha \leq \alpha_{\rm cr}$, where we normalize the variance in such a way that the term $\sum_{\boldsymbol{r}} r^2 H_{\boldsymbol{r}}^2$ in Eq.~(3) does not diverge.
In Fig.~\ref{fig:DeltaX2}{\bf c} we show an additional case with $\alpha > \alpha_{\rm cr}$, similar to Fig.~1{\bf b} of the main text, where no normalization is required.
Interestingly, the transition to the CME happens at $t\sim 1/\gamma$ independently of the hopping range $\alpha$.

\subsection{Moments of the distribution}\label{sec:SImoments}

The moments of the distribution $n_{\boldsymbol{j}}(t)$ can be obtained from the derivatives of the generating function $K$. This can be easily understood from its definition
\begin{equation}
K(\boldsymbol{q},t)=\sum_{\boldsymbol{j}} n_{\boldsymbol{j}}(t)e^{i\boldsymbol{q}\cdot \boldsymbol{j}}.
\end{equation}
Indeed, on a lattice of dimension $d$, we have that
\begin{equation}
    \frac{\partial K}{\partial q_\alpha}(\boldsymbol{0},t)=i\sum_{\boldsymbol{j}} j_\alpha n_{\boldsymbol{j}}(t)=i \langle j_\alpha \rangle, \qquad (\alpha=1,\dots,d)
\end{equation}
\begin{equation}
    \frac{\partial^2 K}{\partial q_\alpha \partial q_\beta} (\boldsymbol{0},t)=-\sum_{\boldsymbol{j}} j_\alpha j_\beta n_{\boldsymbol{j}}(t)=-\langle j_\alpha j_\beta \rangle \qquad (\alpha,\beta=1,\dots,d).
\end{equation}
Therefore, we have that
\begin{equation}
    \nabla_{\boldsymbol{q}} K(\boldsymbol{0},t)=i\langle \boldsymbol{j} \rangle,
\end{equation}
and 
\begin{equation}
    \Delta_{\boldsymbol{q}} K(\boldsymbol{0},t)=-\langle |\boldsymbol{j}|^2 \rangle,
\end{equation}
and similarly, considering higher order derivatives we can compute higher order moments. 

As an example, let us compute explicitly the first two moments of the distribution. The derivatives of $K$ can be evaluated starting from Eq.~(8)
of the main text, yielding
\begin{equation}
\langle \boldsymbol{j} \rangle=\boldsymbol{0},
\end{equation}
\begin{equation}
\label{eq:variance}
\langle |\boldsymbol{j}|^2 \rangle= \mathcal{A}_{2\alpha-2,d}(\boldsymbol{0}) t=2D_\alpha t.
\end{equation}

\subsection{The critical exponent}\label{sec:crit_exp}

In \eqref{eq:variance} can we see that the variance of the distribution depends explicitly on the exponent of the power-law hopping $\alpha$ and on the dimensionality $d$ of the lattice through the function $\mathcal{A}$. In particular, we have that 
\begin{equation}
\label{eq:multizeta}
    \mathcal{A}_{s,d}(\boldsymbol{0})=\kappa \sum_{\boldsymbol{r}\neq\boldsymbol{0}}|\boldsymbol{r}|^{-s}.
\end{equation}
Since the general term appearing in the series in \eqref{eq:multizeta} is a positive and decreasing function of its argument, then we may study its convergence by studying the convergence of the associated integral. In particular, we would have to check the convergence of 
\begin{equation}
   \int_{1}^\infty r^{-s+d-1}dr\propto\big[r^{-s+d}\big]_{1}^\infty,
\end{equation}
which is convergent for $s>d$. Therefore, comparing this result with \eqref{eq:variance}, we have that the variance is finite if and only if 
\begin{equation}
    \alpha>\alpha_{\rm cr}=\frac{d+2}{2}.
\end{equation}

\subsection{Analytical properties of \texorpdfstring{$\mathcal{A}_{2\alpha,d}(\boldsymbol{q})$}{A₂ₐ(q)}}\label{sec:A2alphad}

As stated in the main text, in order to establish the analytical properties of $n_{\boldsymbol{j}}(t)$ for different values of $\alpha$,
we have to study the behaviour of the function $\mathcal{A}_{2\alpha,d}$ around $\boldsymbol{q}=\boldsymbol{0}$. 

For $d=1$ we have that 
\begin{equation}
    \mathcal{A}_{2\alpha,1}(q)=2\kappa\Re(\Li_{2\alpha}(e^{iq})),
\end{equation}
where $\Li_\beta(z)$ denotes the polylogarithm function of order $\beta$ and argument $z$. 

\noindent For $\alpha\neq\frac{1}{2},1,\frac{3}{2},2,\dots$ we can use the expansion about $q=0$ given in \cite{nist}
\begin{equation}
    \label{eq:poly1}
    \mathcal{A}_{2\alpha,1}(q)=-C_{\alpha}|q|^{2\alpha-1}+2\kappa\sum_{j=0}^{\infty}\zeta_{2\alpha-2j}(-1)^j\frac{q^{2j}}{(2j)!},
\end{equation}
with $C_\alpha=-2\kappa\Gamma(1-2\alpha)\sin(\alpha \pi)$, $\Gamma(z)=\int_0^\infty t^{z-1}e^{-z}dz$ being the gamma function and $\zeta_s=\sum_{k=1}^\infty k^{-s}$ the Riemann zeta function.

\noindent For all the other values of $\alpha$, i.e those corresponding to $2\alpha\in\mathbb{N}$, we can derive an expansion about $q=0$ by exploiting some properties of the polylogarithm functions. In particular, considering that
\begin{equation}
    \partial_z \Li_\beta(z)=z^{-1}\Li_{\beta-1}(z),
\end{equation}
and introducing the function $\mathcal{G}_\beta(q)=2i\kappa\Im(\Li_\beta(e^{iq}))$, we can see that
\begin{equation}
    \begin{cases}
    \partial_q\mathcal{A}_{\beta,1}(q)=i \mathcal{G}_{\beta-1}(q) \\
    \partial_q \mathcal{G}_\beta(q)=i \mathcal{A}_{\beta-1,1}(q) \\
    \end{cases}, 
\end{equation}
\begin{equation}
\label{eq:Diff}
    \Rightarrow \;\;\;\partial^2_q \mathcal{A}_{\beta,1}(q)=-\mathcal{A}_{\beta-2,1}(q).
\end{equation}
The boundary conditions for these equations are given by the properties of the functions $\mathcal{G}$ and $\mathcal{A}$ at $q=0$, which can be easily understood by their definition:
\begin{equation}
\label{eq:init}
\mathcal{A}_{\beta,1}(0)=2\kappa\zeta_\beta, \;\; \partial_q\mathcal{A}_{\beta,1}(0)=i\mathcal{G}_{\beta-1}(0)=0~.
\end{equation}
The starting point will be the expression of $\mathcal{A}$ and $\mathcal{G}$ for $\beta=1$, which are given by 
\begin{equation}
\label{eq:uno}
    \begin{cases}
    \mathcal{A}_{1,1}(q)=-2\kappa\log 2 - 2\kappa\log \big|\sin\big(\frac{q}{2}\big)\big|\\
    \mathcal{G}_1(q)=-i\kappa q+i\pi\kappa \sigma(q)
    \end{cases},
\end{equation}
where we denoted the sign function of $q$ by $\sigma(q)$. Let us start our analysis from $\beta=2\alpha$ with $\alpha\in\mathbb{N}$. The following identities will be useful for our procedure:
\begin{equation}
    \int \sigma(q)q^{2j}dq=\frac{|q|^{2j+1}}{2j+1}, \;\;\;\;\;\; \int |q|^{2j-1}=\sigma(q)\frac{q^{2j}}{2j}.
\end{equation}
With them in mind, it is easy to derive $\mathcal{A}_{2,1}$ upon integration of $\mathcal{G}_1$:
\begin{equation}
    \label{eq:a21}
    \mathcal{A}_{2,1}(q)=\kappa\left(\frac{q^2}{2}-\pi|q| + 2\zeta_2\right).
\end{equation}
Similarly, integrating $\mathcal{A}_{2,1}$ twice we get 
\begin{equation*}
    \mathcal{A}_{4,1}(q)=\kappa\left(-\frac{q^4}{4!}+\pi\frac{|q|^3}{3!}-2\zeta_2\frac{q^2}{2!}+2\zeta_4\right).
\end{equation*}
Therefore, it sounds reasonable to assume that we have
\begin{align}
    \label{eq:even}
    \mathcal{A}_{2\alpha,1}(q)=&(-1)^\alpha \frac{\pi}{(2\alpha-1)!}\kappa|q|^{2\alpha-1} 
    +2\kappa\sum_{j=0}^{\alpha}(-1)^j\zeta_{2\alpha-2j}\frac{q^{2j}}{(2j)!}, 
\end{align}
$\forall \alpha \in \mathbb{N}$, with $\zeta_0=-\frac{1}{2}$. This cleary holds for $\alpha=1$, and as well for $\alpha+1$, as we can easily check upon substitution of \eqref{eq:even} in \eqref{eq:Diff} and integrating, yielding to
\begin{align*}
    \mathcal{A}_{2\alpha+2,1}(q)=&(-1)^{\alpha+1} \frac{\pi}{(2\alpha+1)!}\kappa|q|^{2\alpha+1} 
    +2\kappa\sum_{j=0}^{\alpha+1}(-1)^j\zeta_{2\alpha+2-2j}\frac{q^{2j}}{(2j)!}, 
\end{align*}
which proves the validity of \eqref{eq:even} by induction.

\noindent Finally, let us study the $\mathcal{A}_{\beta,1}$ functions with odd integer index, i.e. for $\beta=2s-1$, for $s\in\mathbb{N}$. Of course, it would be wonderful to proceed as for the case of even indices, but this is hard to do since already $\mathcal{A}_{1,1}$ can't be integrated in terms of elementary functions. Therefore, we will limit ourselves to studying the behaviour of these functions close to $q=0$, where they present a logarithmic non-analyticity. In particular, we can see that
\begin{equation}
    \mathcal{A}_{1,1}(q)\approx -\kappa\log q^2.
\end{equation}
Upon integration, we thus get 
\begin{equation}
    \label{eq:tre}
    \mathcal{A}_{3,1}(q)\approx \kappa\left(\frac{q^2}{2}\log q^2+2\zeta_3 -\frac{3}{2}q^2\right),
\end{equation}
\begin{equation}
    \mathcal{A}_{5,1}(q)\approx \kappa\left(-\frac{q^4}{4!}\log q^2+2\zeta_5 -\zeta_3 q^2\right) + O(q^4),
\end{equation}
and, more in general, we have that
\begin{align}
    \label{eq:odd}
    \mathcal{A}_{2s+1,1}(q)\approx &\kappa\left[(-1)^{s+1}\frac{q^{2s}}{(2s)!}\log q^2 
    +2\zeta_{2s+1}-\zeta_{2s-1}q^2\right] + O(q^4),
\end{align}
for $s=2,3,4,\dots$. 
The results in \eqref{eq:poly1},\eqref{eq:even},\eqref{eq:odd} fully characterize the behaviour of $\mathcal{A}$ close to the origin in one dimension, for any value of the exponent $\alpha$.

Determining the behaviour of $\mathcal{A}_{2\alpha,d}$ for $d>1$ is a bit more involved. In order to understand its properties, we will introduce a fixed, small but finite, lattice constant $\lambda$, and approximate the sum appearing in the definition of $\mathcal{A}$ with an integral, i.e. $\sum_{\boldsymbol{r}\neq\boldsymbol{0}}\approx\lambda^{2\alpha-d}\int_0^{2\pi}d\theta\big(\prod_{k=1}^{d-2}\int_0^\pi \sin^{d-1-k}\phi_k d\phi_k \big)\int_\lambda^\infty r^{d-1} dr$. In this way, upon further changing variable to $z=\frac{r}{\lambda}$, we obtain that
\begin{equation}
    \label{eq:intAd}
    \mathcal{A}_{2\alpha,d}\approx\nu_d\kappa\int_1^\infty dz \int_0^\pi d\phi_1 z^{d-2\alpha-1}\sin^{d-2}\phi_1 e^{-iqz\cos(\phi_1)},
\end{equation}
where $q=|\boldsymbol{q}|$ and $\nu_d=2\pi^{\frac{d-1}{2}}/\Gamma\big(\frac{d-1}{2}\big)$ is a constant which only depends on the dimension $d$ of the space. The integral in \eqref{eq:intAd} has to be computed differently for $d=2$ and $d>2$, but after the calculation we are able to write the following expansion about $\boldsymbol{q}=\boldsymbol{0}$ for $\mathcal{A}$ $\forall d \ge 2$:
\begin{equation}
    \label{eq:series1}
    \mathcal{A}_{2\alpha,d}(\boldsymbol{q})\approx\mathcal{A}_{2\alpha,d}(\boldsymbol{0})+\pi^{\frac{d}{2}}2^{d-2\alpha}\frac{\Gamma\big(\frac{d}{2}-\alpha\big)}{\Gamma\big(\alpha\big)}\kappa|\boldsymbol{q}|^{2\alpha-d},
\end{equation}
for $\alpha\le\alpha_{\rm cr}$, and
\begin{align}
    \label{eq:series2}
    \mathcal{A}_{2\alpha,d}(\boldsymbol{q})\approx\mathcal{A}_{2\alpha,d}(\boldsymbol{0})-\frac{\mathcal{A}_{2\alpha-2,d}(\boldsymbol{0})}{2}|\boldsymbol{q}|^2
    +\pi^{\frac{d}{2}}2^{d-2\alpha}\frac{\Gamma\big(\frac{d}{2}-\alpha\big)}{\Gamma\big(\alpha\big)}\kappa|\boldsymbol{q}|^{2\alpha-d},
\end{align}
for $\alpha>\alpha_{\rm cr}$.

\subsection{Asymptotic properties of \texorpdfstring{$\mathbf{n_j(t)}$: $\mathbf{d=1}$}{pⱼ(t): d=1}}\label{sec:asymp1}

Let us start by considering $\alpha < \alpha_{\rm cr}=\frac{3}{2}$. In this case, for $\alpha\neq 1$, we can
take the inverse Fourier transform of \eqref{eq:poly1}
obtaining, to leading order in $q$,
\begin{equation*}
    n_j(t)\approx\frac{1}{2\pi}\int_{-\pi}^\pi e^{-ijq}e^{- tC_\alpha |q|^{2\alpha-1}}dq.
\end{equation*}
Changing variable to $y=jq$ we thus get 
\begin{equation*}
    n_j(t)\approx\frac{1}{2\pi j}\int_{-j\pi}^{j\pi} e^{-iy}e^{-tC_\alpha |y|^{2\alpha-1}/n^{2\alpha-1}}dy.
\end{equation*}
Expanding the $j-$dependent exponential in series we therefore have
\begin{equation*}
    n_j(t)\approx\sum_{l=0}^\infty \frac{(-1)^{l}}{2\pi j^{2\alpha l +1-l}}\frac{(tC_\alpha)^l}{l!}\int_{-j\pi}^{j\pi} |y|^{2\alpha l-l}e^{-iy}dy.
\end{equation*}
For large values of $j$ we can approximate the region of integration with the whole real axis, and to leading order in $j$ we hence obtain that
\begin{equation}
    \label{eq:PLscaling}
    n_j(t)\approx -\frac{t}{j^{2\alpha}}\frac{C_\alpha}{2\pi}\int_{-\infty}^\infty|y|^{2\alpha -1}e^{-iy}dy=\frac{\kappa t}{j^{2\alpha}}. 
\end{equation}
The case $\alpha=1$ can be solved exactly since we have an exact expression for $\mathcal{A}_{2,1}$ in \eqref{eq:a21}. In this case we have that
\begin{align*}
    n_j(t)=&\frac{e^{-\kappa t\pi^2/2}}{\pi}\int_0^\pi \cos(jq) e^{\kappa t(q-\pi)^2/2}dq 
    =
    \frac{e^{-\kappa t\pi^2/2}}{\pi}(-1)^j \int_0^\pi \cos(jy) e^{\kappa t y^2/2}dy, 
\end{align*}
\begin{equation}
    \Rightarrow n_j(t)=\frac{1}{\sqrt{2\kappa t\pi^2}}\bigg[D_F\bigg(\frac{ij+\pi \kappa t}{\sqrt{2\kappa t}}\bigg)-D_F\bigg(\frac{ij-\pi \kappa t}{\sqrt{2\kappa t}}\bigg)\bigg],
\end{equation}
where $D_F(z)=e^{-z^2}\int_0^z e^{u^2}du$ is the Dawson integral. One can check that the asymptotic behaviour for large values of $j$ is 
\begin{equation}
    n_j(t)\approx \frac{\kappa t}{j^2}, 
\end{equation}
which agrees with \eqref{eq:PLscaling}.

\noindent Next, we shall consider the case $\alpha=\frac{3}{2}$, corresponding to the critical point of the model for $d=1$. In such case we find, using the same logic as above, that to leading order
\begin{align}
    n_j(t)&\approx\frac{1}{2\pi}\int_{-\pi}^\pi e^{-ijq} e^{\frac{\kappa t}{2} q^2 \log q^2}dq 
    \approx \frac{\kappa t}{2\pi j^3}\int_{-\infty}^\infty y^2 e^{-iy}\log|y|dy\approx\frac{\kappa t}{j^3},
\end{align}
which again agrees with the power-law tail predicted in \eqref{eq:PLscaling}.

Above the critical point, when $\alpha$ is neither integer or half-integer, we again
take the inverse Fourier transform of \eqref{eq:poly1}.
This time, the leading order in $q$ brings a quadratic Gaussian contribution, while the power with fractional exponent will produce the power-law tail of the distribution: 
\begin{equation*}
    n_j(t)\approx\frac{1}{2\pi}\int_{-\pi}^\pi e^{-ijq}e^{-tC_\alpha |q|^{2\alpha-1}-\kappa t\zeta_{2\alpha-2}q^2 }dq.
\end{equation*}
Changing variable to $y=jq$ and expanding the exponential with the faster decaying exponent for large values of $n$, we get, to leading order, that
\begin{align*}
n_j(t)\approx &\frac{1}{2\pi j}\int_{-\infty}^{\infty} e^{-iy-D_\alpha t y^2 j^{-2}}dy 
- \frac{tC_\alpha}{2\pi j^{2\alpha}}\int_{-\infty}^{\infty}|y|^{2\alpha-1}e^{-iy}dy,   
\end{align*}
where $D_\alpha=\kappa \zeta_{2\alpha-2}$. The above steps give the following asymptotic result for $n_j$:
\begin{equation}
    \label{eq:PLDscaling}
    n_j(t)\approx\frac{e^{-\frac{j^2}{4D_\alpha t}}}{\sqrt{4\pi D_\alpha t}}+\frac{\kappa t}{j^{2\alpha}}. 
\end{equation}
The case of integer $\alpha\in\mathbb{N}$ is analogous since, as we can see in \eqref{eq:even}, the type of singularity at $q=0$ is the same as in the case we have just treated. Hence, we would find once again \eqref{eq:PLDscaling}. 

\noindent Finally, let us consider the case $\alpha$ half integer. In such case, we find that
\begin{equation*}
    n_j(t)\approx\frac{1}{2\pi}\int_{-\pi}^\pi e^{-ijq} e^{\frac{(-1)^{\alpha+\frac{1}{2}}\kappa t}{(2\alpha-1)!} q^{2\alpha-1} \log q^2-D_\alpha tq^2}dq.
\end{equation*}
Therefore, we have that to leading order in $n$
\begin{align*}
    n_j(t)\approx &\frac{1}{2\pi j}\int_{-\infty}^{\infty}e^{-iy-D_\alpha t y^2 j^{-2}}dy 
    + \frac{(-1)^{\alpha+\frac{1}{2}}\kappa t}{\Gamma(2\alpha)\pi j^{2\alpha}}\int_{-\infty}^{\infty}y^{2\alpha-1}e^{-iy}\log|y|dy,   
\end{align*}
resulting once again in 
\begin{equation*}
    n_j(t)\approx\frac{e^{-\frac{j^2}{4D_\alpha t}}}{\sqrt{4\pi D_\alpha t}}+\frac{\kappa t}{j^{2\alpha}},
\end{equation*}
i.e. again \eqref{eq:PLDscaling}.

\subsection{Asymptotic properties of \texorpdfstring{$\mathbf{n_j(t)}$: $\mathbf{d>1}$}{pⱼ(t): d>1}}\label{sec:asympdgt1}

The procedure for the higher dimensional case is a straightforward generalization of what was done in $d=1$, using the power-law expansions \eqref{eq:series1},\eqref{eq:series2}. The final result is the same behaviour observed in one dimension, i.e.

\begin{equation}
\label{eq:PLsummary}
    n_{\boldsymbol{\boldsymbol{j}}}(t)\approx
    \begin{cases}
        \displaystyle \frac{\kappa t}{|\boldsymbol{j}|^{2\alpha}} & \displaystyle \alpha\le\frac{d+2}{2} \\
        \displaystyle \frac{e^{-\frac{|\boldsymbol{j}|^2}{4D_\alpha t}}}{(4\pi D_\alpha t)^{d/2}}+\frac{\kappa t}{|\boldsymbol{j}|^{2\alpha}} & \displaystyle \alpha>\frac{d+2}{2}.
    \end{cases}
\end{equation}

\begin{figure}[!tb]
    \centering
    \includegraphics[width=0.45\textwidth]{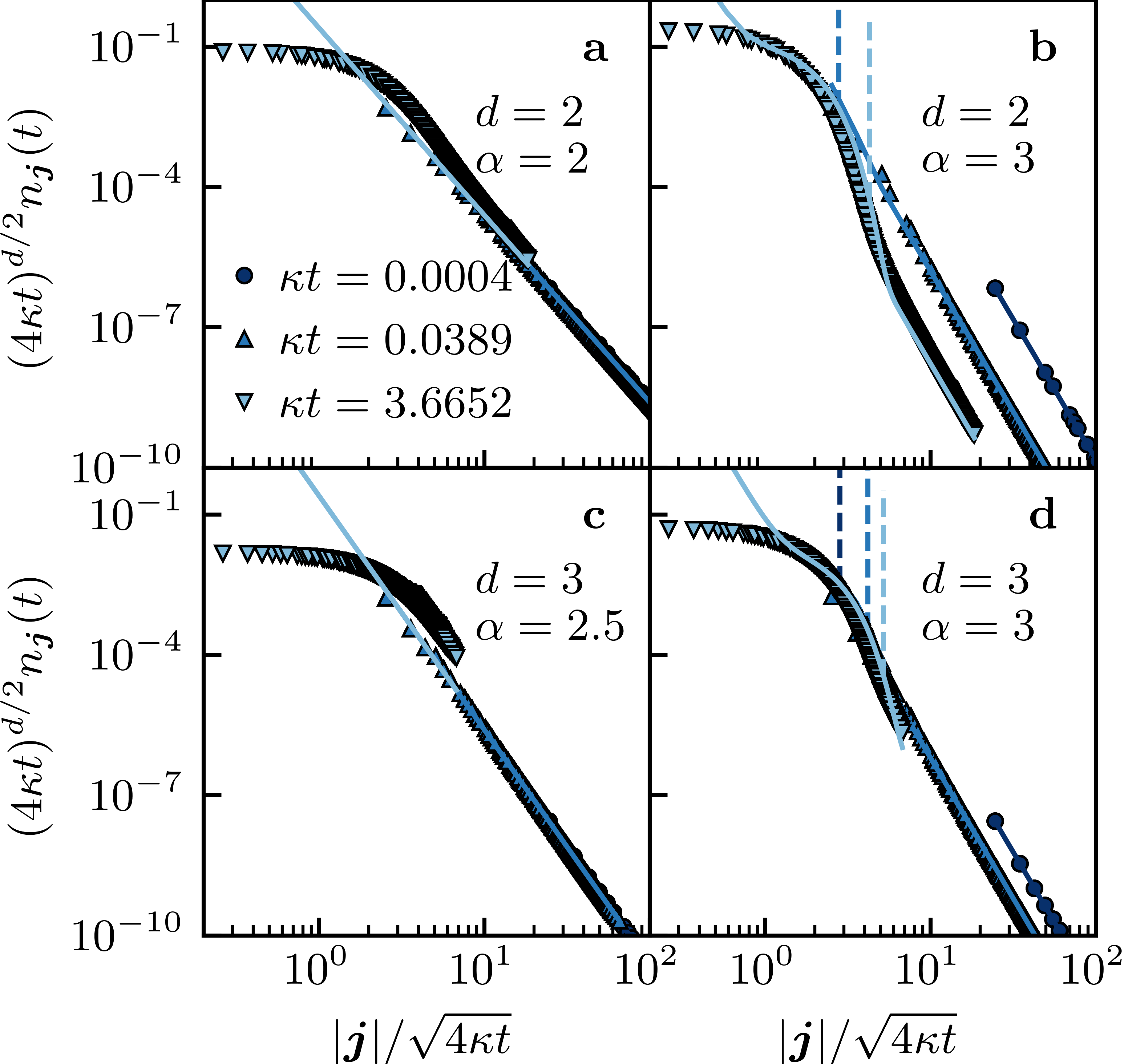}
    \caption{Re-scaled excitation profile at different times $t$ obtained from the CME~(7) for arrays of dimension $d=2$ ({\bf a,b} with $N=100^2$) and $d=3$ ({\bf c,d} with $N=30^3$, respectively. In all panels $\gamma=10J$ and the continuous lines correspond to Eq.~(10) of the main text. The vertical dotted lines correspond to the function $\xi_{\alpha,t}$ entering Eq.~(10b) of the main text. The initial excitation is set on an edge.}
    \label{fig:DeltaX3}
\end{figure}

\subsection{Length scale for the crossover between the Gaussian and the power-law profiles}

Here we derive, for $\alpha>\alpha_{\rm cr}$, the length scale $\xi_{\alpha,t}$ beyond which the excitation profile changes from Gaussian to power-law. This length scale is determined by the crossing point between the Gaussian and the power-law profile, i.e.
\begin{equation}
    \frac{e^{-\xi_{\alpha,t}^2/4D_\alpha t}}{(4\pi D_\alpha t)^{d/2}} = \frac{\kappa t}{\xi_{\alpha,t}^{2\alpha}}~.
\end{equation}
We define here $x=\xi_{\alpha,t}^2/4D_\alpha t$ and $b=(4D_\alpha t)^{\alpha_{\rm cr} - \alpha}\pi^{d/2} \kappa/4D_\alpha$ so that we have
\begin{equation}
    x^\alpha e^{-x} = b~,
\end{equation}
which can be solved by means of the Lambert $W$ function. Specifically, our problem is equivalent to $y e^y = -b^{1/\alpha}/\alpha$, where $y=-x/\alpha$. The Lambert $W$ function is defined by its inverse, i.e. $W(y e^y)=y$. In our case, since $y$ is negative, there are two possible solutions (the ``0'' and the ``$-1$'' branch of the $W$ function), corresponding to the two intersections between a power law and a Gaussian. Since we are interested in the largest length scale, we choose the ``$-1$'' branch, so that we have $y=W_{-1}(-b^{1/\alpha}/\alpha)$. Upon substituting the definition of $y$ we have $x=-\alpha W_{-1}(-b^{1/\alpha}/\alpha)$, and recovering the definitions of $x$ and $b$, we have finally
\begin{equation}\label{lambertW}
    \xi_{\alpha,t} = \sqrt{ -4\alpha D_\alpha t \,W_{-1}\left(-\frac{1}{\alpha} \left(\frac{\pi^{d/2} \kappa}{4D_\alpha} (4D_\alpha t)^{\alpha_{\rm cr} - \alpha}\right)^{1/\alpha} \right) }~.
\end{equation}
Note that $\xi_{\alpha,t}$ exists only within the domain of the $W$ function, that is for
\begin{equation}
    t \geq t_{\rm cr} = \frac{1}{4D_\alpha} \left[ \frac{\pi^{d/2}\kappa e^\alpha}{4D_\alpha \alpha^\alpha} \right]^{1/(\alpha-\alpha_{\rm cr})}~.
\end{equation}

Eq.~\eqref{lambertW} is exact, but not very illuminating about the dependence of $\xi_{\alpha,t}$ on $\alpha$ and $t$. Anyway, for very large $t$, the argument of the $W$ function tends to $0^-$, and it can be approximated as $W_{-1}(y)\approx \log(-y)$. In this way we obtain
\begin{equation}
    \xi_{\alpha,t} \approx \sqrt{ 4 D_\alpha t \,\log \left(  \frac{4\alpha^\alpha D_\alpha}{\pi^{d/2} \kappa} (4D_\alpha t)^{\alpha - \alpha_{\rm cr}}\right) }~,
\end{equation}
which is the length scale reported in the main text.

\subsection{Diffusion-enhancement in the case of F\"orster energy transfer}

 The square of the exciton diffusion length (in units of the lattice parameter) corresponds to the variance of the exciton distribution:
\begin{equation}
 L^{2}_{\alpha}=\langle |\boldsymbol{j}|^2 \rangle(\tau)=\sum_{{\boldsymbol r}\neq\boldsymbol{0}} r^{2} \kappa (r) \tau,
 \end{equation}
 with $\tau$ the exciton lifetime, and 
 \begin{equation}
 \kappa (r)\propto \sum_{{\boldsymbol r}\neq\boldsymbol{0}} \vert H_{\boldsymbol r} \vert^{2} \int \! d\omega F_{\rm D} (\omega) \sigma_{\rm A} (\omega)
 \end{equation}
 the transfer rate between the sites $\boldsymbol{j}$ (donor) and $\boldsymbol{j}+\boldsymbol{r}$ (acceptor), separated by the distance $r$. Here, $H_{\boldsymbol r}=\bra{G}S^-_{\boldsymbol{j}} H S^+_{\boldsymbol{j}+{\boldsymbol r}}\ket{G}$ are the matrix elements of the Hamiltonian Eq.~(2) connecting the two states involved in the energy transfer, and $\ket{G}$ is the ground state with all spins down. The functions 
\begin{align}
F_{\rm D} (\omega) = \sigma_{\rm A} (\omega) = \frac{1}{\pi} \frac{\gamma}{(\omega-\omega_{0})^{2} + \gamma^{2}}
\end{align}
entering the overlap integral denote the normalized donor emission spectrum and acceptor absorption spectrum. Here we assume that all spins have the same energy $\omega_{0}$, or equivalently that the energy difference between two sites is negligible compared to the FWHM $\gamma$ (corresponding to the dephasing rate). One thus finds $L^{2}_{\alpha} \propto \kappa \tau \mathcal{A}_{2\alpha-2,d}(\boldsymbol{0})/2$, with $\mathcal{A}_{2\alpha,d}(\boldsymbol{0})=\sum_{\boldsymbol{r}\neq\boldsymbol{0}}r^{-2\alpha}$ and $\kappa=2 J^{2}/\gamma$, as defined in the main text. Our diffusion length $L_{\alpha}$ including the contribution of the long-range tail can be compared with that of the standard theory of diffusion assuming nearest-neighbor hopping ($\alpha=\infty$). The latter reads $L^{2}_{\infty} \propto \kappa \tau d$ (with $d$ the dimension). For F\"orster energy transfer ($\alpha=3$) at play in nanocrystal films, the ratio between the square diffusion lengths in the long-range and nearest-neighbor cases is
\begin{equation}
\frac{L^{2}_{3}}{L^{2}_{\infty}} \approx \begin{cases} 2.8 \; \textrm{for} \; d=3 \\
1.5 \; \textrm{for} \; d=2 \end{cases}
\end{equation}
We emphasize that this factor $\sim 2$ depending on the dimensionality of the energy transfer (the latter can neither be considered as fully 2D nor fully 3D in Refs.~\cite{Giovanni2021,Mork2014}) accounts for part of the discrepancy between the exciton diffusion length measured experimentally and the values predicted by standard (nearest-neighbor) diffusion theory applied to F\"orster energy transfer~\cite{Giovanni2021,Mork2014}.

\section{The many-particle problem}

\subsection{Derivation of the effective Liouvillian}\label{sec:MBpert}

Here we look at the many-body case starting from the long-range Hamiltonian (2) and following the idea of \cite{Cai}, which is to study the large-dephasing limit of the model through a second-order perturbative analysis, deriving an effective Liouvillian $\hat{\mathcal{L}}_{\rm eff}$ in the limit $\gamma \gg J$. Therefore, we split the original Liouvillian, Eq.~(1), into two contributions, an unperturbed term $\hat{\mathcal{L}}_0\rho=\gamma \sum_{\boldsymbol{j}}(S_{\boldsymbol{j}}^z\rho S_{\boldsymbol{j}}^z-\rho/4)$, and a perturbation $\hat{\mathcal{L}}_1\rho=-i[H,\rho]$. The steady states of $\hat{\mathcal{L}}_0$ are given by $\ket{\boldsymbol{\sigma}}\bra{\boldsymbol{\sigma}}$ where the $\ket{\boldsymbol{\sigma}}$ are eigenstates of the $\{S_{\boldsymbol{j}}^z\}$ operators, i.e. $S_{\boldsymbol{j}}^z\ket{\boldsymbol{\sigma}}=s_{\boldsymbol{j}}^z\ket{\boldsymbol{\sigma}}$, with $s_{\boldsymbol{j}}^z=\pm \frac{1}{2}$. Following Ref.~\cite{Cai}, the effective Liouvillian projected onto the diagonal subspace generated by the $\ket{\boldsymbol{\sigma}}\bra{\boldsymbol{\sigma}}$ reads
\begin{equation}
\label{eq:Leff}
    \hat{\mathcal{L}}_{\rm eff}=\hat{\mathcal{P}}\hat{\mathcal{L}}_1\frac{1}{\lambda_0-\hat{\mathcal{L}}_0}\hat{\mathcal{L}}_1\hat{\mathcal{P}},
\end{equation}
where $\hat{\mathcal{P}}$ is the projector onto this subspace.

At this point, we need to evaluate $\hat{\mathcal{L}}_0(\hat{\mathcal{L}}_1\ket{\boldsymbol{\sigma}}\bra{\boldsymbol{\sigma}})$.
Using the notation $h_{\boldsymbol{j}\boldsymbol{r}}=S^+_{\boldsymbol{j}} S^-_{\boldsymbol{j}+\boldsymbol{r}}+S^-_{\boldsymbol{j}} S^+_{\boldsymbol{j}+\boldsymbol{r}}$, we have that
\begin{equation}
    \label{eq:L1}
    \hat{\mathcal{L}}_1\ket{\boldsymbol{\sigma}}\bra{\boldsymbol{\sigma}}=-i\frac{J}{2}\sum_{\boldsymbol{j}; \boldsymbol{r}\neq\boldsymbol{0}}[h_{\boldsymbol{j}\boldsymbol{r}},\ket{\boldsymbol{\sigma}}\bra{\boldsymbol{\sigma}}]|\boldsymbol{r}|^{-\alpha}.
\end{equation}
The commutator in \eqref{eq:L1} is given by 
\begin{equation}
    [h_{\boldsymbol{j}\boldsymbol{r}},\ket{\boldsymbol{\sigma}}\bra{\boldsymbol{\sigma}}]=\delta_{s_{\boldsymbol{j}+\boldsymbol{r}}^z,-s_{\boldsymbol{j}}^z}(\ket{\boldsymbol{\sigma}',\boldsymbol{j}\boldsymbol{r}}\bra{\boldsymbol{\sigma}}-\;h.\,c.),
\end{equation}
where $\ket{\boldsymbol{\sigma}',\boldsymbol{j}\boldsymbol{r}}=\ket{\ldots -s_{\boldsymbol{j}}^z \ldots -s_{\boldsymbol{j}+\boldsymbol{r}}^z \ldots}$. Substituting into \eqref{eq:L1}, we obtain
\begin{equation}
    \label{eq:L1_1}
    \hat{\mathcal{L}}_1\ket{\boldsymbol{\sigma}}\bra{\boldsymbol{\sigma}}=-i\frac{J}{2}\sum_{\boldsymbol{j}; \boldsymbol{r}\neq\boldsymbol{0}}\frac{\delta_{s_{\boldsymbol{j}+\boldsymbol{r}}^z,-s_{\boldsymbol{j}}^z}}{|\boldsymbol{r}|^{\alpha}}(\ket{\boldsymbol{\sigma}',\boldsymbol{j}\boldsymbol{r}}\bra{\boldsymbol{\sigma}}-\;h.\,c.).
\end{equation}
Here, if we apply $\hat{\mathcal{L}}_0$ to an element of the sum of \eqref{eq:L1_1}, we obtain
\begin{align}
    \hat{\mathcal{L}}_0\ket{\boldsymbol{\sigma}',\boldsymbol{j}\boldsymbol{r}}\bra{\boldsymbol{\sigma}}=\frac{\gamma}{4}\sum_{\boldsymbol{j}}(4 s_{\boldsymbol{j}}'^z s_{\boldsymbol{j}}^z-1)\ket{\boldsymbol{\sigma}',\boldsymbol{j}\boldsymbol{r}}\bra{\boldsymbol{\sigma}}
    =-\gamma \ket{\boldsymbol{\sigma}',\boldsymbol{j}\boldsymbol{r}}\bra{\boldsymbol{\sigma}}, 
\end{align}
therefore yielding to
\begin{equation}
    \hat{\mathcal{L}}_0(\hat{\mathcal{L}}_1\ket{\boldsymbol{\sigma}}\bra{\boldsymbol{\sigma}})=-\gamma\hat{\mathcal{L}}_1\ket{\boldsymbol{\sigma}}\bra{\boldsymbol{\sigma}}.
\end{equation}

Now, considering also that $\lambda_0=0$ for the steady states, the effective Liouvillian \eqref{eq:Leff} is reduced to
\begin{align}
    \hat{\mathcal{L}}_{\rm eff}\ket{\boldsymbol{\sigma}}\bra{\boldsymbol{\sigma}}&=\frac{1}{\gamma}\hat{\mathcal{P}}\hat{\mathcal{L}}_1\hat{\mathcal{L}}_1\ket{\boldsymbol{\sigma}}\bra{\boldsymbol{\sigma}}
    = -\frac{1}{\gamma}\hat{\mathcal{P}}[H,[H,\ket{\boldsymbol{\sigma}}\bra{\boldsymbol{\sigma}}]] \nonumber \\
    &=-\frac{J^2}{4\gamma}\sum_{\boldsymbol{j};\boldsymbol{r}\neq \boldsymbol{0}}\sum_{\boldsymbol{m};\boldsymbol{r}'\neq \boldsymbol{0}}|\boldsymbol{r}|^{-\alpha}|\boldsymbol{r}'|^{-\alpha}\hat{\mathcal{P}}(h_{\boldsymbol{j}\boldsymbol{r}}h_{\boldsymbol{m}\boldsymbol{r}'}\ket{\boldsymbol{\sigma}}\bra{\boldsymbol{\sigma}}+\ket{\boldsymbol{\sigma}}\bra{\boldsymbol{\sigma}}h_{\boldsymbol{j}\boldsymbol{r}}h_{\boldsymbol{m}\boldsymbol{r}'}-2h_{\boldsymbol{j}\boldsymbol{r}}\ket{\boldsymbol{\sigma}}\bra{\boldsymbol{\sigma}}h_{\boldsymbol{m}\boldsymbol{r}'}),
\end{align}
with $h_{\boldsymbol{j}\boldsymbol{r}}=S^+_{\boldsymbol{j}} S^-_{\boldsymbol{j}+\boldsymbol{r}}+S^-_{\boldsymbol{j}} S^+_{\boldsymbol{j}+\boldsymbol{r}}$.
On applying the projector $\hat{\mathcal{P}}$, the only non-vanishing terms have $h_{\boldsymbol{m}\boldsymbol{r}'}=h_{\boldsymbol{j}\boldsymbol{r}}$, and for each $h_{\boldsymbol{j}\boldsymbol{r}}$ we have two such terms: $(\boldsymbol{m}=\boldsymbol{j};\boldsymbol{r}'=\boldsymbol{r})$ and $(\boldsymbol{m}=\boldsymbol{j}+\boldsymbol{r};\boldsymbol{r}'=-\boldsymbol{r})$. Therefore we drop the sum over $\boldsymbol{m};\boldsymbol{r}'$ and multiply by a factor of 2, obtaining
\begin{equation}
    \hat{\mathcal{L}}_{\rm eff}\ket{\boldsymbol{\sigma}}\bra{\boldsymbol{\sigma}}=-\frac{J^2}{2\gamma}\sum_{\boldsymbol{j};\boldsymbol{r}\neq \boldsymbol{0}}|\boldsymbol{r}|^{-2\alpha}(h^2_{\boldsymbol{j}\boldsymbol{r}}\ket{\boldsymbol{\sigma}}\bra{\boldsymbol{\sigma}}+\ket{\boldsymbol{\sigma}}\bra{\boldsymbol{\sigma}}h^2_{\boldsymbol{j}\boldsymbol{r}}-2h_{\boldsymbol{j}\boldsymbol{r}}\ket{\boldsymbol{\sigma}}\bra{\boldsymbol{\sigma}}h_{\boldsymbol{j}\boldsymbol{r}}).
\end{equation}
Following Ref.~\cite{Cai}, we note that $h^2_{\boldsymbol{j}\boldsymbol{r}}=2\left(\frac{1}{4}-S_{\boldsymbol{j}}^zS_{\boldsymbol{j}+\boldsymbol{r}}^z\right)$. Therefore, we obtain the CME (5) with generator (6) reported in the main text.


\subsection{Occupation probability for the 1D symmetric exclusion process with long-jumps}
\label{ap:MB}

To give some quantitative analysis of the many-particle case we decided to consider the case of a one dimensional lattice of $N$ sites indexed by $\{j\}_{-N/2}^{N/2-1}$, with open boundary conditions. We take as initial state the configuration where the $N/2$ sites on the left of the origin are all occupied, while the remaining sites are empty. We are interested in understanding how the occupation probability $n_j(t)$ at site $j$ evolves.
In particular, considering the flow in and out of each lattice site, we can write the following discrete time evolution for $n_j(t)$:
\begin{equation}
    n_j(t+\Delta t)-n_j(t)=[1-n_j(t)]\sum_{l=-N/2, l\neq j}^{N/2-1}[n_l(t)n^{(sp)}_{l-j}(\Delta t)]-n_j(t)\sum_{l=-N/2, l\neq j}^{N/2-1}[1-n_l(t)]n^{(sp)}_{l-j}(\Delta t),
\end{equation}

\begin{equation}
    \label{eq:occ1}
    \Rightarrow n_j(t+\Delta t)-n_j(t)=\sum_{l=-N/2, l\neq j}^{N/2-1}[n_{l}(t)-n_j(t)]n^{(sp)}_{l-j}(\Delta t),
\end{equation}
where $n^{(sp)}_{j}(\Delta t)$ denotes the single-particle hopping probabilities on the lattice. To estimate them, we can look at the short time solution of their master equation, which we derived in the first section of this manuscript
\begin{equation}
    \label{eq:prob}
    \dot{n}^{(sp)}_j=\kappa\sum_{r=-N/2-j, r\neq 0}^{N/2-1-j}(n^{(sp)}_{j+r}-n^{(sp)}_j)|r|^{-2\alpha},
\end{equation}
with initial condition $n^{(sp)}_j(0)=\delta_{j,0}$. Therefore, for short times and $j\neq 0 $ we get
\begin{equation}
    \label{eq:ST}
    n^{(sp)}_j(\Delta t)\approx\frac{\kappa \Delta t}{j^{2\alpha}}. 
\end{equation}
Inserting \eqref{eq:ST} in \eqref{eq:occ1} we thus obtain
\begin{equation}
\label{eq:occ}
    \dot{n}_j=\kappa\sum_{l=-N/2, l\neq j}^{N/2-1}(n_{l}-n_j)|l-j|^{-2\alpha},
\end{equation}
which we will have to solve with the initial condition $n_j(0)=1-\Theta[j]$, where $\Theta$ is the discrete Heaviside step function. Since we used a short-time approximation to derive \eqref{eq:occ}, it makes sense to solve it in the same approximation. Therefore, for $j\ge0$ we would have
\begin{equation}
    n_j(\Delta t)\approx\kappa \Delta t \sum_{l=-N/2,l\neq j}^{N/2-1}(1-\Theta[l])|l-j|^{-2\alpha},
\end{equation}
\begin{equation}
\label{occR}
    \Rightarrow n_j(\Delta t)\approx\kappa \Delta t \sum_{l=-N/2}^{-1}|l-j|^{-2\alpha}=\kappa \Delta t \sum_{r=j+1}^{N/2+j}|r|^{-2\alpha}.
\end{equation}
This has a really simple physical interpretation. Indeed, it amounts to say that the probability that a site $j\gg 0$ is occupied after a short time $t$ corresponds to the independent probabilities that at least one particle has jumped to that site starting from the step-function initial configuration. 
Similarly, for $j<0$ we have
\begin{equation}
    n_j(\Delta t)\approx1-\kappa \Delta t \sum_{l=-N/2,l\neq j}^{N/2-1}\Theta[l]|l-j|^{-2\alpha},
\end{equation}
\begin{equation}
    \Rightarrow n_j(\Delta t)\approx1-\kappa \Delta t \sum_{l=0}^{N/2-1}|l-j|^{-2\alpha},
\end{equation}
and hence
\begin{equation}
\label{eq:occL}
    \Rightarrow n_j(\Delta t)\approx1-\kappa \Delta t \sum_{r=|j|}^{N/2-1+|j|}|r|^{-2\alpha}.
\end{equation}
Again, this has a simple physical explanation, i.e. after a short time $t$ the occupation of a site on the left of the chain is only influenced by the escape probability of the particle that initially was at that position. 

Putting \eqref{occR} and \eqref{eq:occL} together we thus get the occupation profile for short time:
\begin{equation}
    n_j(t)=\Theta[j] \kappa t\sum_{r=j+1}^{N/2+j}|r|^{-2\alpha} + (1-\Theta[j])\big[1-\kappa t \sum_{r=|j|}^{N/2-1+|j|}|r|^{-2\alpha}\big].
\end{equation}

To get some insight about the large-time regime we can instead look at the stationary solution of \eqref{eq:occ1}. In this case we would have 
\begin{equation}
    \sum_{l=-N/2, l\neq j}^{N/2-1}[n_{l}-n_j]n^{(sp)}_{l-j}=0,
\end{equation}
which can be solved by setting $n_j=c\in\mathbb{R}$ $\forall j$. The constant is determined by the normalization condition which implies the conservation of the number of particles, i.e. $\sum_l n_l=\frac{N}{2}$. Therefore, we have that for large $t$ the stationary solution is the flat profile
\begin{equation}
    n_j=1/2 \;\;\; \forall j.
\end{equation}

\section{The quantum case: weak-dephasing regime}
\label{sec:PT}

Here we consider the QME [Eq.~(1) in the main text] in the single-exciton case, for weak dephasing and $d=1$. We rewrite the Hamiltonian [Eq.~(2) in the main text] as
\begin{equation}
    H = \sum_{1 \leq i<j \leq N} h_{i,j} (S_i^+ S_j^- + h.c.),
\end{equation}
where $h_{i,j}=J[|i-j|^\alpha + (N-|i-j|)^\alpha](1-\delta_{i,j})$ is the single-particle Hamiltonian with periodic boundary conditions.
The evolution of the two-point correlation functions $G_{j,m}=\Tr(\rho S^+_jS^-_m)$ in the single-particle regime is described by Eqs.~(\ref{eq:Coherences}-\ref{eq:population}), that we rewrite in matrix form as
\begin{equation}
    \label{eq:QMEG}
    \dot{G}=i[h^T,G]-\gamma [G-\text{diag}(G)]~.
\end{equation}
In order to solve Eq.~\eqref{eq:QMEG}, here we follow the approach of \cite{eisler2011crossover}. Thanks to the time-linearity of Eq.~\eqref{eq:QMEG}, we can solve it in terms of the eigenvalues and eigenvectors of the time-evolution operator ($E_{q,k}$ and $A^{q,k}$ respectively, with some labels $q,k$ that we characterize below). We substitute $A_{j,m}^{q,k}(t)=A_{j,m}^{q,k}e^{-E_{q,k} t}$ into Eq.~\eqref{eq:QMEG}, leaving us with the eigenvalue equation
\begin{equation}
    \label{eq:masterA}
    -E_{q,k} A_{j,m}^{q,k}=i\sum_{l=0}^{N-1}[h_{l,j}A_{l,m}^{q,k}-A_{j,l}^{q,k}h_{m,l}]-\gamma(1-\delta_{j,m}) A_{j,m}^{q,k}. 
\end{equation}
At this point we exploit translational invariance to write $A_{j,m}^{q,k}=e^{iqj}A_{0,m-j}^{q,k}$, where $q=\frac{2\pi}{N}j$, $j=0,\dots,N-1$. Substituting this in Eq.~\eqref{eq:masterA} we thus have 
\begin{equation}
    \label{eq:eigv}
    E_{q,k} A_{0,m-j}^{q,k}=-i\sum_{l=0}^{N-1}\big[h_{l,j}A_{0,m-l}^{q,k}e^{iq(l-j)}-A_{0,l-j}^{q,k}h_{m,l}\big]+\gamma(1-\delta_{0,m-j}) A_{0,m-j}^{q,k}. 
\end{equation}
This equation can be written in matrix form as
\begin{equation}
\label{eq:eig}
(C_q+\gamma X)\vec{A}^{q,k}=E_{q,k}\vec{A}^{q,k}.
\end{equation}
Here, $\vec{A}^{q,k}$ are vectors of size $N$ with components $A^{q,k}_{0,m}$ ($m=0,\dots,N-1$), while $C_q$ are $N\times N$ circulant matrices with elements
\begin{equation}
\label{eq:Cq}
(C_q)_{m,j}=
i\big[1-e^{iq(m-j)}\big]h_{m,j} \qquad (m,j=0,\cdots,N-1)
\end{equation}
and $X$ is a diagonal matrix with elements $X_{m,m} = 1-\delta_{0,m}$ (where $m=0,\dots,N-1$).
Therefore, we have to solve $N$ independent eigenvalue problems of the type of Eq.~\eqref{eq:eig} (one for each value of $q$), where each provides $N$ eigenvectors and eigenvalues, $\vec{A}^{q,k}$ and $E_{q,k}$, for some labeling index $k=0,\dots,N-1$.
Then, given an initial condition $G(0)$ for the two-point correlation matrix, the time-evolved $G(t)$ is
\begin{equation}
    \label{eq:evolved}
    G_{j,m}(t)=\sum_{q,k}\Tr{[(A^{q,k})^{-1} G(0)]}A^{q,k}_{j,m}e^{-E_{q,k}t}.
\end{equation}


Since an analytical solution of the eigenvalue problem in Eq.~\eqref{eq:eig} for $q\neq0$ is non-trivial, in the following we focus on the small-dephasing regime, and we treat $\gamma X$ as a perturbation of $C_q$.


\subsection{Small dephasing: perturbation theory}

For the unperturbed problem ($\gamma=0$) the normalized eigenstates $(\vec{A}^{q,k})^{(0)}$ and the corresponding eigenvalues $E_{q,k}^{(0)}$ ($k=0,\frac{2\pi}{N},\dots,2\pi \frac{N-1}{N}$) of the circulant matrices $C_q$ are 
\begin{equation}
(A^{q,k}_{0,m})^{(0)}=\frac{e^{imk}}{\sqrt{N}},
\end{equation}
\begin{equation}
\label{eq:eigenvalues}
E_{q,k}^{(0)} = \sum_{m=0}^{N-1} (C_q)_{0,m} e^{imk}.
\end{equation}

From Eq.~\eqref{eq:Cq} one can see that $C_q$ are anti-hermitian matrices. Therefore, all the eigenvalues in \eqref{eq:eigenvalues} are purely imaginary conjugated pairs, except one which is equal to zero~\footnote{For even values of $N$ exact degeneracies appear in the unperturbed spectrum, therefore the results obtained here assume an odd number of sites in the lattice.}. In order to use standard perturbation theory formulas for the eigenvalues of hermitian operators we multiply both sides of Eq.~\eqref{eq:eig} by the imaginary unit, so
that, up to correction of order $\gamma^{4}$, we have
\begin{equation}
\label{eq:pt3}
E_{q,k}=E^{(0)}_{q,k}+\gamma \delta^{(1)}_{q,k}-i\gamma^2\delta^{(2)}_{q,k}+\gamma^3\delta^{(3)}_{q,k} + o(\gamma^4),
\end{equation}
with the following real coefficients:
\begin{equation}
\delta^{(1)}_{q,k}=\frac{N-1}{N},
\end{equation}
\begin{equation}
\delta^{(2)}_{q,k}=\frac{i}{N^2}\sum_{p\neq k}\frac{1}{E^{(0)}_{q,p}-E^{(0)}_{q,k}},
\end{equation}
\begin{equation}
\delta^{(3)}_{q,k}=\frac{1}{N^3}\sum_{p\neq k}\sum_{s\neq k,p}\frac{1}{(E^{(0)}_{q,k}-E^{(0)}_{q,p})(E^{(0)}_{q,k}-E^{(0)}_{q,s})}.
\end{equation}

The first order of the perturbation series gives a constant contribution to the real part of the eigenvalues, whose fluctuations are captured by the third-order term. The second-order term captures the fluctuations of the imaginary part of the eigenvalues. These fluctuations are small, as one can see from 
Fig.~\ref{fig:DecayRate}: for large system sizes the smallest real part of the eigenvalues predicted by \eqref{eq:pt3} (green symbols) converges to $\gamma$, and the convergence does not depend on the value of the hopping rate $\alpha$. This means that the correction obtained from the first-order perturbation theory is effectively dominant. Therefore, for weak dephasing, we can keep just the leading order corrections for the eigenvectors,
\begin{equation}
    A_{0,m}^{q,k}=(A_{0,m}^{q,k})^{(0)}-\frac{\gamma}{N}\sum_{p\neq k}\frac{(A_{0,m}^{q,p})^{(0)}}{E^{(0)}_{q,k}-E^{(0)}_{q,p}},
\end{equation}
and eigenvalues,
\begin{equation}
    E_{q,k}=E^{(0)}_{q,k}+\gamma \frac{N-1}{N},
\end{equation}
which, when plugged into \eqref{eq:evolved}, would imply that $G_{j,m}$ relaxes to its stationary state with a fixed rate $\sim\gamma$.

\subsection{Exact diagonalization}

\begin{figure}[!tb]
    \centering
    \includegraphics[width=.45\columnwidth]{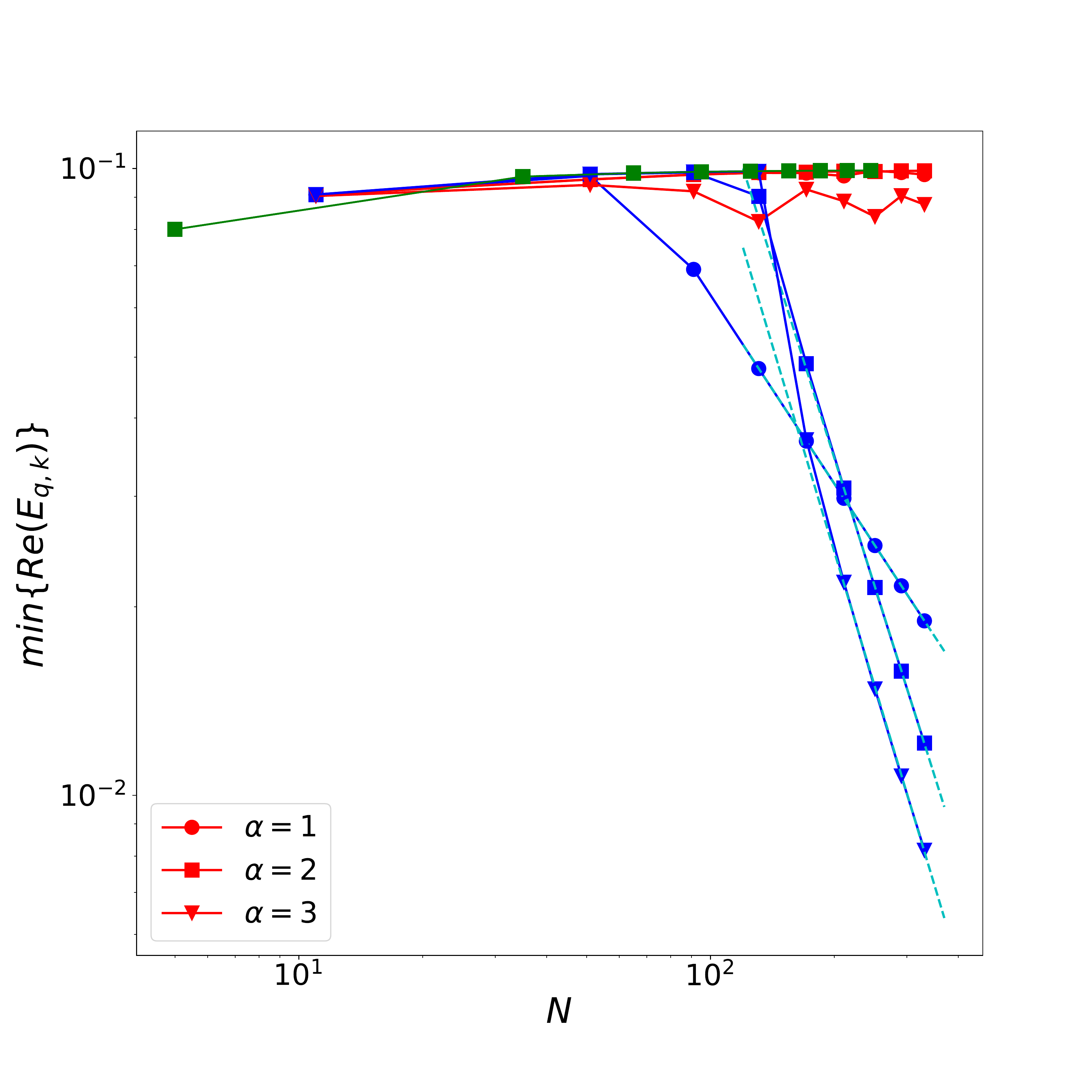}
    \caption{Scaling of the smallest real part for complex Liouvillian eigenvalues (red) and real ones (blue) with the system size. The green symbols show the results obtained for the real eigenvalues using perturbation theory, while the cyan dashed lines are power-law fits ($\sim 1/N$ for $\alpha=1$ and $\sim 1/N^2$ for $\alpha=2,3$). Data are taken for $\alpha=1$, $2$, $3$, $J=1$ and $\gamma=0.1$. 
    }
    \label{fig:DecayRate}
\end{figure} 

\begin{figure}[!tb]
    \centering
    \includegraphics[width=.48\columnwidth]{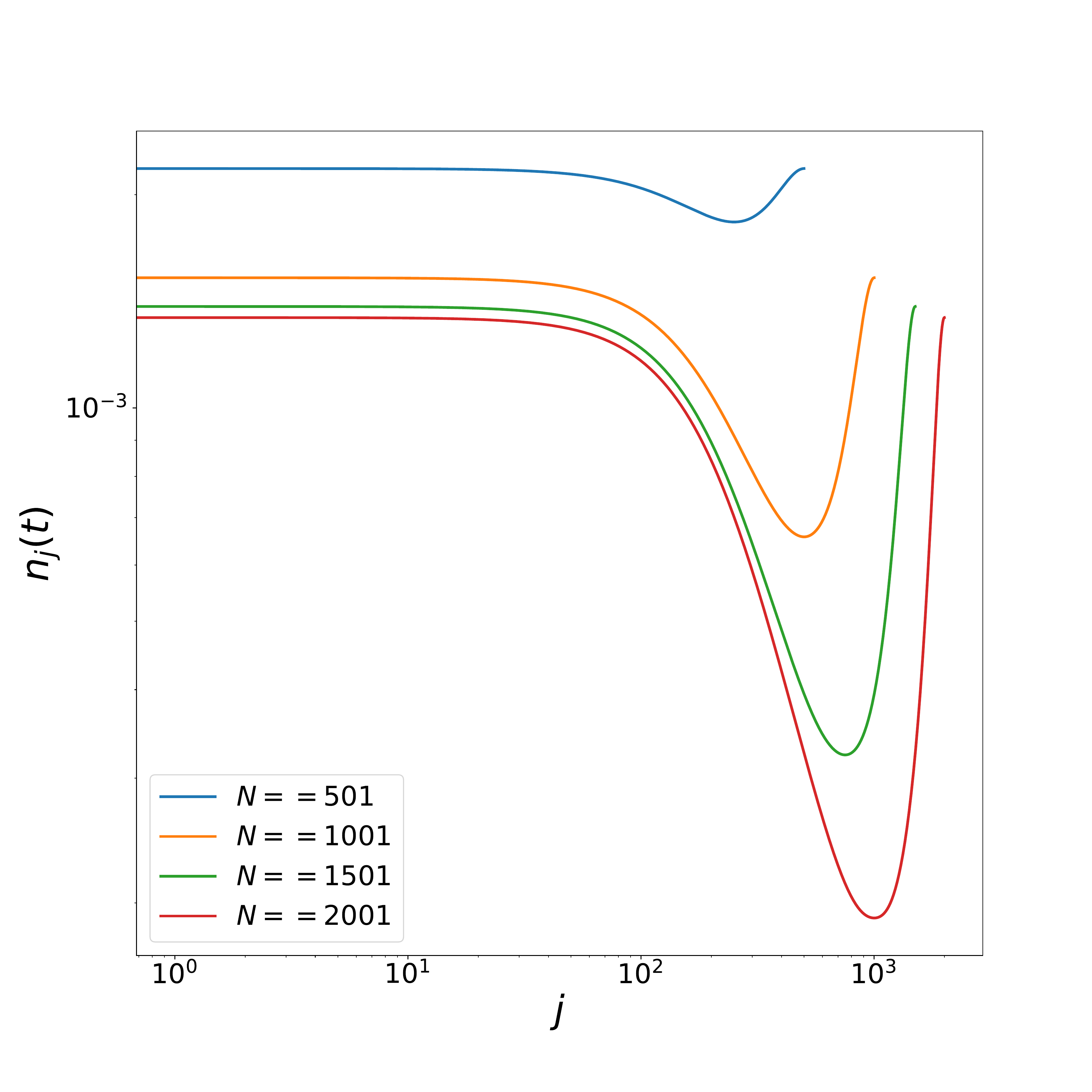}
    \includegraphics[width=.48\columnwidth]{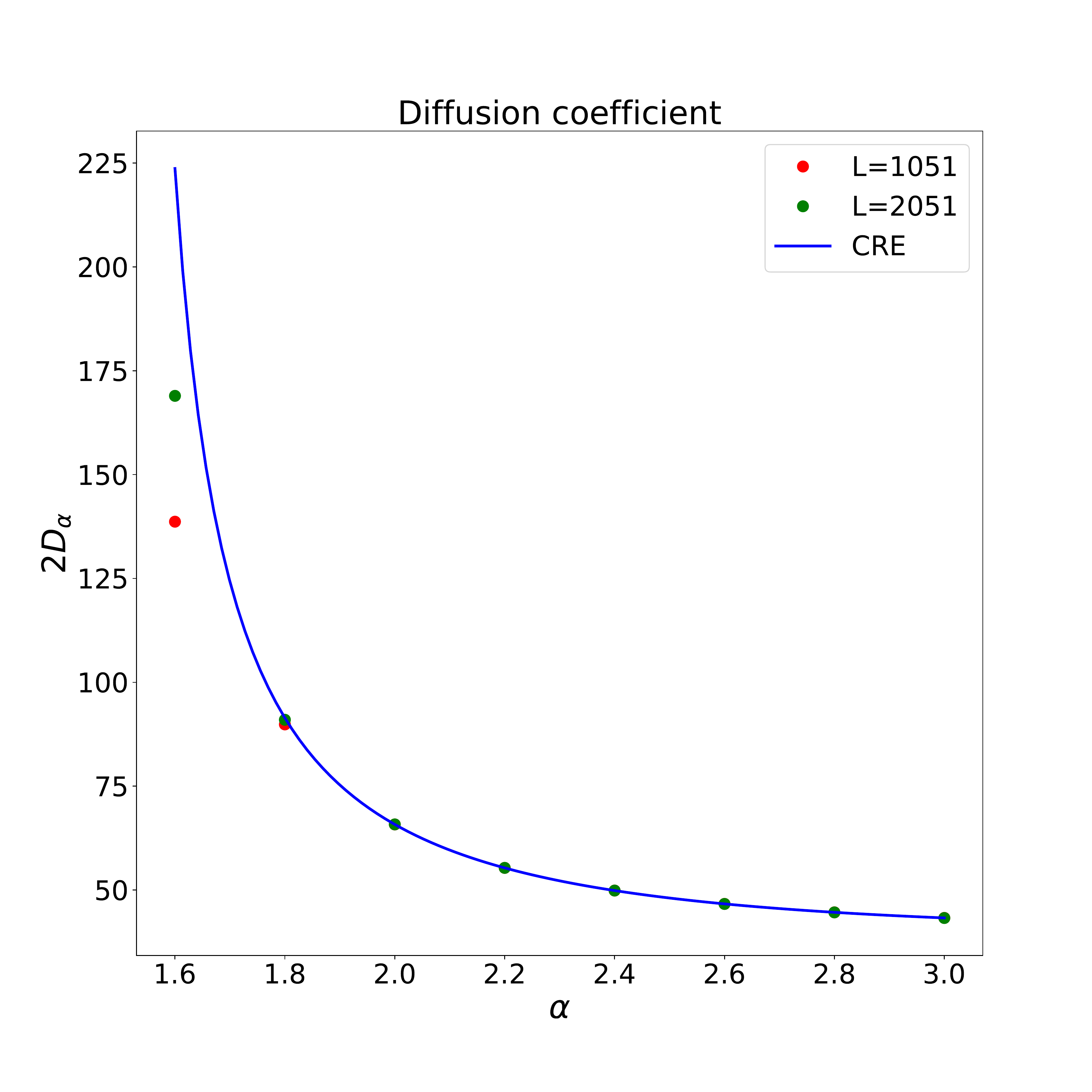}
    
    \caption{Left: exciton density profile for $\alpha=J=1$, $\gamma=0.1$ and $t=250$ for different system sizes $N$. Right: diffusion constant determined by fitting the variance of the exciton density {\it vs.} time, $\braket{|\boldsymbol{j}|^2}=2D_\alpha t$, for $J=1$, $\gamma=0.1$.
    }
    \label{fig:transport}
\end{figure}

An exact diagonalization approach reveals that the real part of the vast majority of the eigenvalues are indeed $\sim \gamma$, independent of $\alpha$ (see red symbols in Fig.~\ref{fig:DecayRate}). However, we find also that there are $N-1$ real eigenvalues which, for large system sizes,
acquire a decay rate smaller than $\gamma$ (see blue symbols in Fig.~\ref{fig:DecayRate}). This is not captured by the perturbation theory, and thus the long-time dynamic, which is dominated by these small eigenvalues, is non-perturbative. Therefore, in the long-time limit we can restrict our analysis to these slow-decaying terms. The results obtained with exact numerics are shown in Fig.~\ref{fig:transport} (left panel). 
The logarithmic plot shows the emergence of a power-law tail in the exciton density profile $n_j=G_{j,j}$ before reaching a minimum at $j=N/2$. There, the profile bends and the density increases because of the periodic boundary conditions that we have employed.
Comparing the amplitudes of the populations and coherences obtained in the long-time limit we found that the populations are at least one order of magnitude larger than the coherences. Therefore, this justifies an adiabatic elimination of the coherences in Eq.~\eqref{eq:QMEG} similarly to the large dephasing regime (see Sec.~\ref{sec:QME-CRE}) which would lead us to the classical master equation \eqref{eq:RE1}. Indeed, as one can see in the right panel in Fig.~\ref{fig:transport}, the diffusion constant obtained numerically in the weak-dephasing regime (symbols) agrees quite well with that predicted by the CME (continuous line), and the agreement improves as we increase the system size.

\bibliographystyle{apsrev4-2}
\bibliography{ref.bib}